\documentclass[apj,twocolumn,twocolappendix,numberedappendix]{openjournal}
\usepackage{newtxtext,newtxmath,enumitem,multirow}
\usepackage[utf8]{inputenc}
\usepackage[T1]{fontenc}
\usepackage[breaklinks, colorlinks,allcolors=blue,citecolor=blue,urlcolor=blue]{hyperref}
\usepackage{orcidlink}
\usepackage{ctable}
\usepackage[normalem]{ulem}
\pdfoutput=1

\usepackage{makecell}

\newcommand{\etal}{et al.}

\newcommand{\lya}{Ly$\alpha$}
\newcommand{\lyas}{Ly$\alpha$'s}
\newcommand{\lyapar}{(Ly$\alpha$)\,}

\newcommand{\lyaemng}{Ly$\alpha$-emitting}
\newcommand{\lyaper}{Ly$\alpha$.}
\newcommand{\lyacom}{Ly$\alpha$,}

\newcommand{\dittoclosing}{---''---}

\newcommand{\NII}{N{\sc ~ii}}
\newcommand{\OII}{O{\sc ~ii}}
\newcommand{\OIII}{O{\sc ~iii}}
\newcommand{\mexp}{\mathrm{exp}}
\newcommand{\mlya}{\mathrm{Ly}\alpha}

\newcommand{\kms}{\mbox{km\ s${^{-1}}$}}
\newcommand{\cmjj}{\mbox{${\rm cm^{-2}}$}}

\shorttitle{Cosmic Noon Spatially Resolved \lya}
\shortauthors{Solhaug \etal}

\begin{document}
\title[Cosmic Noon Spatially Resolved \lya]{Deciphering Spatially Resolved Lyman-alpha Profiles in Reionization Analogs: \\ \vspace{1mm}the Sunburst Arc at Cosmic Noon} 

\author{Erik Solhaug\,\orcidlink{0000-0003-3302-0369}$^{1,2,\star}$}
\author{Hsiao-Wen Chen\,\orcidlink{0000-0001-8813-4182}$^{1,2}$}
\author{Mandy C.\ Chen\,\orcidlink{0000-0002-8739-3163}$^{1,2}$}
\author{Fakhri Zahedy\,\orcidlink{0000-0001-7869-2551}$^{3}$}
\author{Max Gronke\,\orcidlink{0000-0003-2491-060X}$^{4}$}
\author{Magdalena J. Hamel-Bravo\,\orcidlink{0009-0008-5041-4347}$^{5,6}$}
\author{Matthew B.\ Bayliss\,\orcidlink{0000-0003-1074-4807}$^{7}$}
\author{Michael D.\ Gladders\,\orcidlink{0000-0003-1370-5010}$^{1,2}$}
\author{Sebastián López\,\orcidlink{0000-0002-2644-0077}$^{8}$}
\author{Nicolás Tejos\,\orcidlink{0000-0002-1883-4252}$^{9}$}
\vspace{1mm}
\affiliation{$^{1}$Department of Astronomy  \& Astrophysics, The University of Chicago, Chicago, IL 60637 USA}
\affiliation{$^{2}$Kavli Institute for Cosmological Physics, The University of Chicago, Chicago, IL 60637 USA}
\affiliation{$^{3}$Department of Physics, University of North Texas, Denton, TX 76201, USA} 
\affiliation{$^{4}$Max-Planck Institute for Astrophysics, Karl-Schwarzschild-Str. 1, D-85741 Garching, Germany}
\affiliation{$^{5}$Centre for Astrophysics and Supercomputing, Swinburne University of Technology, Hawthorn, Victoria 3122, Australia}
\affiliation{$^{6}$ARC Centre of Excellence for All Sky Astrophysics in 3 Dimensions (ASTRO 3D), Australia}
\affiliation{$^{7}$Department of Physics, University of Cincinnati, Cincinnati, OH 45221, USA}
\affiliation{$^{8}$Departamento de Astronomía,
Universidad de Chile, Casilla 36-D, Santiago, Chile}
\affiliation{$^{9}$Instituto de Física, Pontificia Universidad Católica de Valparaíso, Casilla 4059, Valparaíso, Chile}
\thanks{$^\star$\href{mailto:eriksolhaug@uchicago.edu}{eriksolhaug@uchicago.edu}}


\begin{abstract}
The hydrogen Lyman-alpha (\lya) emission line, the brightest spectral feature of a photoionized gas, is considered an indirect tracer of the escape of Lyman continuum (LyC) photons, particularly when the intergalactic medium is too opaque for direct detection. However, resonant scattering complicates interpreting the empirical properties of \lya\ photons, necessitating radiative transfer simulations to capture their strong coupling with underlying gas kinematics.
In this study, we leverage the exceptional spatial resolution from strong gravitational lensing to investigate the connection between \lya\ line profiles and LyC leakage on scales of a few 100 pc in the Sunburst Arc galaxy at $z\!\sim\!2.37$.
New optical echelle spectra obtained using Magellan MIKE show that both the LyC leaking and non-leaking regions exhibit a classic double-peak \lya\ feature with an enhanced red peak, indicating outflows at multiple locations in the galaxy. Both regions also show a central Gaussian peak atop the double peaks, indicating directly escaped \lya\ photons independent of LyC leakage. We introduce a machine learning-based method for emulating \lya\ simulations to quantify intrinsic dynamics ($\sigma_{\mathrm{int}}$), neutral hydrogen column density ($N_{\mathrm{HI}}$), outflow velocity ($\varv_{\mathrm{exp}}$), and effective temperature ($T$) across continuous parameter spaces. By comparing the spatially and spectrally resolved \lya\ lines in Sunburst, we argue that the directly escaped \lya\ photons originate in a volume-filling, warm ionized medium spanning $\sim\!1$ kpc, while the LyC leakage is confined to regions of $\lesssim\!200$ pc. 
These sub-kpc variations in \lya\ profiles highlight the complexity of interpreting integrated properties in the presence of inhomogeneous mixtures of gas and young stars, emphasizing the need for spatially and spectrally resolved observations of distant galaxies.

\end{abstract}


\maketitle



\section{Introduction}
\label{sec:intro}

To understand whether star-forming galaxies could have reionized the Universe, it is necessary to find links between observables and the escape of ionizing Lyman continuum (LyC) photons.  Direct observations of the LyC escape at high redshifts are hindered by the opaque intergalactic medium (IGM).
Instead, several positive correlations between the escape fraction of LyC photons, $f_{\mathrm{esc, LyC}}$, and the hydrogen \lya\ line, the brightest spectral feature of a photoionized gas, have been suggested, including: \lya\ equivalent width \citep{Verhamme_2017, Fletcher_2019, Flury_2022b, Saldana-Lopez_2022, Pahl_2023}, the flux ratio between the \lya\ trough and the median stellar continuum \citep[$f_{\mathrm{trough}} / f_{\mathrm{cont}}$,][]{Gazagnes_2020}, the fraction of central \lya\ flux within $\pm 100 \, \mathrm{km} \, \mathrm{s}^{-1}$ of the rest-frame velocity \citep[$f_{\mathrm{cen}}$,][]{Naidu_2022}, and \lya\ full width at half maximum \citep[][]{Kramarenko_2024}. Additionally, observational evidence in nearby green-pea galaxies suggests an inverse correlation between the separation of the blue and red scattered \lya\ peaks ($\varv_{\mathrm{sep}}$) and $f_{\mathrm{esc, LyC}}$ \citep{Verhamme_2017, Izotov_2018c, Gazagnes_2020, Izotov_2021, Flury_2022b, Izotov_2022, Naidu_2022}.

The quest to identify galaxies with non-zero $f_{\mathrm{esc, LyC}}$ (so-called LyC leakers) has been ongoing for over two decades \citep[e.g.,][]{Steidel_2001}. Several dozen sources have been earmarked as potential LyC leakers using indirect spectral or photometric diagnostics, including [\OIII]\,$\lambda\lambda\,4959, 5007$ / [\OII]\,$\lambda\lambda$\,3727, 3729 or [\NII]\,$\lambda$\,6584 / H$\alpha$ ratios, or 
$\varv_{\mathrm{sep}}$ \citep[e.g.,][]{Bergvall_2006,Leitet_2011, Leitet_2013, Borthakur_2014, Mostardi_2015, deBarros_2016, Izotov_2016c, Izotov_2016b, Leitherer_2016, Shapley_2016, Bian_2017, Izotov_2018a, Izotov_2018c, Steidel_2018,  Izotov_2020, Izotov_2021, Izotov_2022}. Yet, only a handful of galaxies beyond $z \geq 2$ have gained widespread acceptance as definitive LyC leakers \citep[e.g.,][]{Vanzella_2012, Dahle_2016, deBarros_2016, Shapley_2016, Vanzella_2016, Vanzella_2018, Marques-Chaves_2021, Marques-Chaves_2022, Marques-Chaves_2024}. While more post-reionization LyC-leaking candidates have been identified based on indirect tracers of LyC escape, spatial resolution limitations have often precluded the detailed examination of small-scale features crucial for deciphering LyC escape mechanisms. The scarcity of LyC leakers in the cosmos underscores the critical need for extracting meaningful statistics from 
galaxies that resemble the first galaxies, often referred to as "reionization analogs," particularly in their spectral characteristics. 

This paper explores the potential of \lya\ as a key indicator of ionizing photon escape, leveraging insights from the resonant properties of \lya\ to shed light on how neutral hydrogen species in the interstellar medium (ISM) affect the observed \lya\ profiles, as well as the possible routes of LyC escape. Motivated by the potential relations connecting \lya\ to the escape of ionizing photons, we carry out a quantitative analysis of the \lya\ profiles of distinct star-forming regions in a reionization analog galaxy.

Our target is the Sunburst Arc at $z=2.37$, a particularly bright gravitationally lensed star-forming galaxy at cosmic noon discovered by \cite{Dahle_2016}. The foreground lensing cluster PSZ1 G311.65-18.48 at $z_{\rm lens}=0.443$ 
was discovered through its Sunyaev-Zel’dovich effect in data from the Planck survey \citep{Planck_2014}. Gravitationally lensed arcs, such as the Sunburst Arc, offer a powerful setting for examining the gaseous medium in galaxies beyond the nearby universe. These giant arcs manifest when massive foreground galaxy clusters distort and magnify background galaxies (see e.g., Figures \ref{fig:spectralocations} and \ref{fig:lens} below). This makes these giant arcs well-suited for probing small-scale (${\sim}50\,\mathrm{pc}$) spatial variations, which have unveiled a dynamic environment of stellar winds and continuous flows of high-column density gas in \lyaemng\ galaxies \citep{Chen_2021}. 

\begin{figure}
 \centering
 \includegraphics[width=0.95\columnwidth]{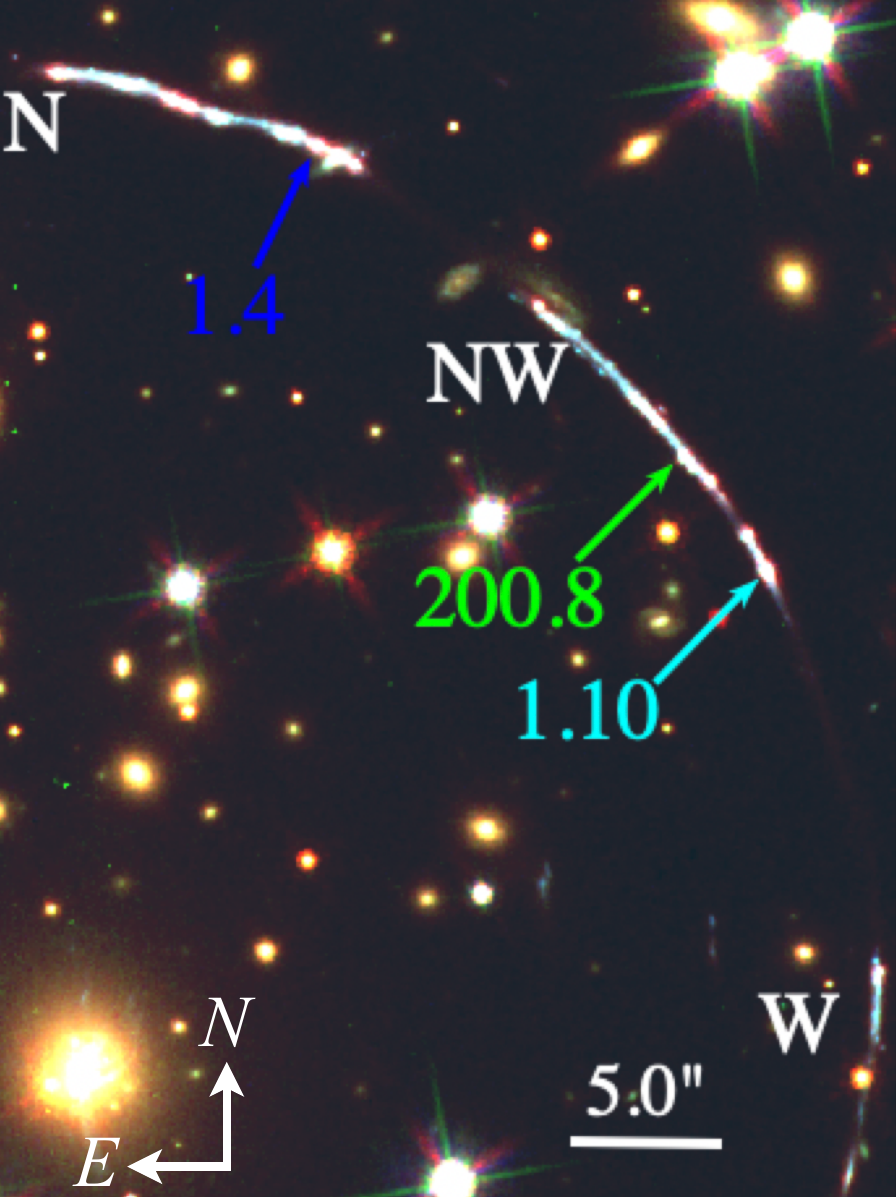}
    \caption{
    Composite image of the field around Sunburst Arc, a star-forming galaxy at $z\!=\!2.37$ gravitationally magnified by a foreground cluster at $z_{\rm lens}\!=\!0.443$.  The image focuses on the northern (N), northwestern (NW), and western (W) portions of the Arc, covering a sky area of $30''\times 40''$.  It is produced using \textit{HST} WFC3 F606W, ACS F814W, and WFC3 F160W images (see \S\ \ref{sec:imagingdata} for details).  The locations of the star-forming knots targeted by the MIKE echelle spectroscopy described in \S\ \ref{sec:echellespectroscopy} are marked with an arrow and their associated image number.  Images 1.4 and 1.10 are two of the multiply-imaged LyC leaker \citep[e.g.,][]{Rivera-Thorsen2019}, while image 200.8 is the single highly-magnified image of a LyC non-leaking region \citep[e.g.,][]{Diego_2022,Sharon2022} 
    }
        \label{fig:spectralocations}
\end{figure}

The Sunburst Arc (Figure \ref{fig:spectralocations}; see also Figure \ref{fig:lens} below) stands out not only for its elevated star formation rate (SFR) but also for its strong LyC leakage from one of its multiply lensed star-forming regions \citep{Rivera-Thorsen2019}. This LyC-leaking region is known to exhibit a triple-peaked \lya\ profile \citep[e.g.,][see also Figure \ref{fig:spectraoverview} below]{Rivera-Thorsen2017, Mainali_2022} which is thought to be comprised of a double-peaked scattered \lya\ component with high $\varv_{\mathrm{sep}}$ and a single non-scattered peak centered on the \lya\ rest frame velocity.

Typically, galaxies with high $\varv_{\mathrm{sep}}$ are not expected to have high $f_{\mathrm{esc, LyC}}$, as the dense neutral gas responsible for producing double peaks also hinders LyC escape. However, ionized cavities or chimneys can facilitate anisotropic escape of LyC while simultaneously scattering \lya\ into double peaks with large $\varv_{\mathrm{sep}}$. The presence of the central peak in the triple-peaked \lya\ profile suggests direct escape of \lya\ photons without scattering, and given that neutral hydrogen is about four orders of magnitude more opaque to \lya\ photons than to ionizing radiation, one would anticipate LyC escape in the presence of a central \lya\ peak.

The Sunburst Arc galaxy thus offers a compelling case to explore the relationship between \lya\ profile morphology and LyC escape. Thanks to strong gravitational lensing, with magnification factors ranging from approximately ten to several hundreds \citep{Rivera-Thorsen2019, Pignataro_2021, Diego_2022, Sharon2022}, individual star-forming regions within the host galaxy can be resolved. This enhanced resolution has, for instance, provided clear evidence of stronger ionized gas outflows in the LyC leaking region compared to non-leaking regions, as evidenced by broad [\OIII]\,$\lambda\,5007$ in the LyC leaker \citep{Mainali_2022}.

In this paper, we leverage the spatial resolution from gravitational lensing and the enhanced spectral resolution of new Magellan MIKE spectra to investigate the observed \lya\ profile and its connection to LyC leakage.  The observations target two spatially resolved star-forming regions within the Sunburst galaxy: one LyC leaking region with an estimated escape fraction of $f_{\mathrm{esc}} \approx 24\%$ and $30\%$ for images 1.4 and 1.10, respectively \citep[][]{Rivera-Thorsen2019}, and one non-leaking region with no detected escape of ionizing photons, for which we estimate an upper limit of $f_{\mathrm{esc}} \lesssim\!0.9\%$ \footnote{We estimate an upper limit of the LyC escape fraction of the non-leaker, $f_{\mathrm{esc}, 200.8}$, using aperture photometry by assuming that the non-leaker shares the same intrinsic spectral shape and line-of-sight IGM opacity as the leaker.  We adopt a circular aperture with a diameter set to equal the FWHM of image 1.10 of the leaker and the non-leaker (image 200.8) in the archival F275W imaging and determine $AB(\mathrm{F275W})_{1.10} \approx 25.6$. Since image 200.8 is a non-detection in F275W, we estimate a 2-$\sigma$ upper limit on its flux by measuring the root-mean-square (RMS) noise in a nearby sky patch and summing the noise in quadrature within the same-sized aperture used for image 1.10, yielding an upper-limit magnitude of $AB(\mathrm{F275W})_{200.8} \approx 29.4$. We estimate the non-ionizing UV apparent magnitudes in the F814W imaging following the same method used for image 1.10 in F275W, finding $AB(\mathrm{F814W})_{1.10} \approx 22.2$ and $AB(\mathrm{F814W})_{200.8} \approx 22.2$ (they have a roughly similar non-ionizing UV magnitude). Recall that at $z=2.37$, the observed F275W and F814W bandpasses correspond to $\approx 800$ and 2400 \AA, respectively, in the rest frame. We then estimate $f_{\mathrm{esc}, 200.8}$, based on $f_{\mathrm{esc}, 1.10} \times 10^{-0.4\,[AB(\mathrm{F275W})_{200.8} - AB(\mathrm{F275W})_{1.10}]} \times 10^{-0.4\,[AB(\mathrm{F814W})_{1.10} - AB(\mathrm{F814W})_{200.8}]} \lesssim 0.9\%$. 
}. As demonstrated below, both the leaking and non-leaking regions exhibit a similar triple-peaked profile, revealing a more complex configuration between dense gas and young stars in distant star-forming regions that may regulate how these ionizing photons escape. Through a quantitative \lya\ profile analysis, we aim to provide new insights into the relationship between the \lya\ line morphology and LyC escape.

\begin{table*}\renewcommand{\arraystretch}{1.5}
	\centering
     \caption{Nomenclature for individual images of star-forming clumps used in the lens models of this work and previous studies.}
	\begin{tabular}{c|c|c|c|c|c}
    \hline
    \hline
    This work & \cite{Pignataro_2021} & \cite{Diego_2022} & \cite{Sharon2022} & ($\mathrm{RA_{J2000}},\, \mathrm{DEC_{J2000}}$) & LyC leaker? \\
    \hline
    \hline
    1.4 & 5.1d & d & 1.4 & (15:50:07.629, $-$78:10:59.60) & yes \\
    \hline
    1.10 & 5.1l & l & 1.10 & (15:50:08.620, $-$78:11:13.63) & yes \\
    \hline
    200.8 & Tr & Tr or "Godzilla" & 4.8 & (15:50:08.436, $-$78:11:09.92) & no \\
    \hline
    \hline
    300.80 and 300.81 or "the pair"\footnote{~These sources are too faint and close to nearby sources to obtain an individual slit spectrum.} & - & Pair & 4.8 & (15:50:08.444, $-$78:11:10.07) & no \\
    \hline
    \hline
	\end{tabular}
     \label{tab:nomenclature}
\end{table*}
\vskip 0.25cm

The structure of this paper is as follows. In Section \ref{sec:data}, we describe the data incorporated in our analysis, including optical echelle spectroscopy and high spatial resolution imaging data.  In Section \ref{sec:lyaspectralanalysis}, we outline our methods for conducting a comprehensive and quantitative analysis of the \lya\ profiles of distinct star-forming regions in Sunburst. In Section \ref{sec:gravitationallensmodel}, we describe our strategy for parametrically modeling the mass distribution of the foreground cluster to reconstruct the source-plane morphology of Sunburst. In Section \ref{sec:results}, we examine the spatial variation of gas properties across $\sim\!1\,\mathrm{kpc}$ based on the observed \lya\ profiles. In Section \ref{sec:discussion}, we discuss the implications on the \lya\ and LyC leakage in reionization analogs, and in Section \ref{sec:summary} we summarize and conclude.  Throughout this paper, we assume a flat cosmology with $H_0\!=\!70\, \mathrm{km}\,\mathrm{s}^{-1}\,\mathrm{Mpc}^{-1}$, $\Omega_{\rm M}\!=\!0.3$, and $\Omega_{\Lambda}\!=\!0.7$.

\section{Data}
\label{sec:data}

A combination of optical echelle spectroscopy and space-based imaging data is included in our study of the Sunburst Arc. High-resolution echelle spectroscopy is needed to resolve the intricate kinematic profiles of the \lya\ line, while space-based imaging data are utilized to enable precise position measurements of the multiply imaged star-forming clumps, or knots, within the source galaxy for constraining the lens model of the foreground galaxy cluster. Here, we describe the acquisitions of these spectral and imaging data.

\subsection{Optical Echelle Spectroscopy}
\label{sec:echellespectroscopy}

High-resolution echelle spectroscopy ($R \approx 29{,}000$) of the knots within the giant lensed arcs in the Sunburst Arc was obtained with the MIKE echelle spectrograph on the Magellan Clay Telescope at Las Campanas Observatory. MIKE is a double-arm high-resolution optical spectrograph. The two arms provide full wavelength coverage from approximately 3350--5000 \AA\ in the blue channel and 4900--9500 \AA\ in the red channel. We performed the observations using a $1.0\!\times\!5.0$ arcsec slit and $2\!\times\!4$ (spatial $\times$ spectral) binning during readout. Wavelength calibrations were performed using a frame of an internal ThAr lamp taken immediately after each exposure and were later corrected to vacuum and heliocentric wavelengths. The observations were challenging due to seeing conditions, which were rarely below $0.7$ arcsec and exceeded $1.0$ arcsec for large parts of the run, combined with cirrus clouds and a high airmass ($>1.5$).

Our spectra cover one pointing in the northern arc (image 1.4) and two pointings in the northwestern arc (images 200.8 and 1.10; see Figure \ref{fig:spectralocations}). The Sunburst Arc has been extensively studied, and various nomenclatures for the multiply lensed images have been developed in parallel with the construction of gravitational lens models. Image nomenclature is typically adopted in the development of a lens model, and since all lens models have their own advantages and differences, there is no universal assignment of image names. We adopt a nomenclature based on that of \cite{Sharon2022} due to the close similarities in our own parametric lens model approach for the Sunburst Arc. To clarify the differences in names between lens models in various works, we summarize the nomenclature in Table \ref{tab:nomenclature}.  Briefly, images 1.4 and 1.10 are two of the multiple  images of the highly-magnified LyC leaking clump in Sunburst \citep[][]{Rivera-Thorsen2019}, while image 200.8 is a bright single image of a non-LyC leaking source at $\lesssim\!1$ kpc in the source plane from the leaker (see \S\ \ref{sec:gravitationallensmodel} below). The image pair near 200.8 was not included as a separate source or independent image constraints but rather as part of Source 4 in \cite{Sharon2022}. Since we include this pair as a distinct source with its own constraints in our lens model (see \S\ \ref{sec:gravitationallensmodel}), we assign the names 300.80 and 300.81 to reflect their origin from the individual Source 300.

Our final coadded spectra comprise exposures of 7.7 hrs (image 1.4), 11.2 hrs (image 200.8), and 9.5 hrs (image 1.10).
This results in a mean signal-to-noise ratio of approximately 16 across our three spectra. The mean line spread function (LSF) for the MIKE spectrograph is determined to have $\mathrm{FWHM} \approx 10\,\mathrm{km}\,\mathrm{s}^{-1}$ based on standard arc lines. For our analysis, we focus on the redshifted \lya\ line from the Sunburst arc, which occurs in MIKE's blue channel.

\begin{figure}
 \centering
 \includegraphics[width=\columnwidth] {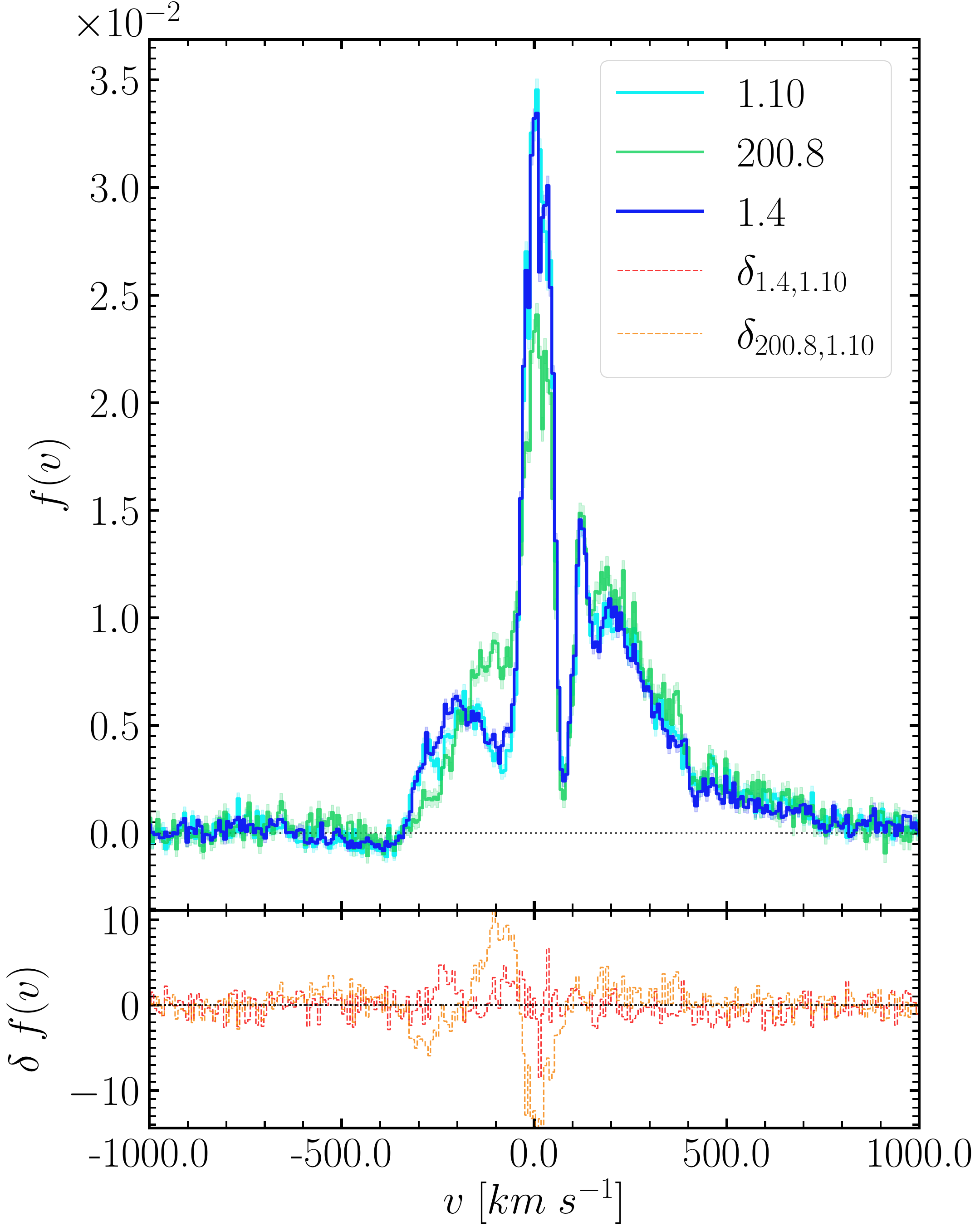}
    \caption{MIKE echelle spectra of the \lya\ signals from 1.4 (blue), 200.8 (green), and 1.10 (cyan) in the Sunburst Arc with associated error spectra displayed as shaded regions around the spectral pixels. Zero velocity corresponds to $z=2.37094$, the fiducial redshift of the LyC leaker \citep[e.g.,][]{Rivera-Thorsen2017}. All spectra have had the continuum subtracted based on a best-fit linear model obtained for the continuum flux over the spectral region of $\pm\,1400$ \kms, excluding the \lya\ line over $\pm\,1000$ \kms.  In addition, the spectral fluxes, $f (\varv)$, are normalized by the total line flux enclosed within the velocity range of $\pm\,1400$ \kms relative to the rest frame of the \lya\ line. The lower panel displays the residuals, $\delta \, f(\varv)$, between each pair of spectra, normalized by their errors summed in quadrature, to emphasize the variation of their \lya\ profiles.  In addition to the classic double-peak \lya\ profile, the spectra are of significantly high quality to reveal small-scale absorption features that are likely due to IGM \lya\ forest absorption.  See \S\ \ref{sec:foregroundabsorbers} for more discussions.
    }
        \label{fig:spectraoverview}
\end{figure}

The velocity profile of the \lya\ line is calibrated to the systemic redshift of $z=2.37094$, determined by \cite{Rivera-Thorsen2017} based on H$\alpha$, H$\beta$, and [\OIII]\,$\lambda\lambda\,4959$, 5007 detected from image 1.10 (see Figure \ref{fig:spectralocations}). 
 We convert our MIKE spectra from observed wavelengths to line-of-sight velocity offsets in the rest frame of \lya\ signals. The subsequent \lya\ profile analysis is limited to the rest frame velocity range of $[-1400, \, +1400] \, \mathrm{km} \, \mathrm{s}^{-1}$.  The final stacked \lya\ profiles of individual knots are presented in Figure \ref{fig:spectraoverview}.

\subsection{Imaging Data}
\label{sec:imagingdata}

Optical images of the Sunburst Arc (PSZ1G311.65-18.48) from the \textit{Hubble Space Telescope (HST)} are retrieved from the {\it HST} archive.  These include data taken with the Wide Field Camera 3 (WFC3) and the F160W and F606W filters (PID: 15377; PI: M. Bayliss), and with the Advanced Camera for Surveys (ACS) and the F814W filter (PID: 15101; PI: H. Dahle). All images are processed using the standard reduction pipeline. We stack these images and register the stacked frames using Gaia astrometry \citep{Gaia_2016, Gaia_2023} of nearby stars in the field. Our composite RGB image (F160W, F814W, F606W) for the field is displayed in Figures \ref{fig:spectralocations} and \ref{fig:lens}.

\section{\lya\ Spectral Analysis}
\label{sec:lyaspectralanalysis}

While the \lya\ line is the strong spectral feature expected of a photoionized gas, interpreting the line signals is challenging due to its resonant scattering nature \citep[e.g.,][]{Dijkstra:2019}.  The large cross section between neutral hydrogen and the \lya\ photons indicates that these photons will likely undergo a series of random walks both in space and in frequency before escaping their host galaxies.  While the observed \lya\ profile encodes the underlying gas kinematics, an accurate interpretation of the spectral feature requires detailed radiative transfer calculations.

In this section, we first compare and demonstrate that the observed \lya\ profiles vary between different images within the Sunburst Arc.  Then we describe the \lya\ radiative transfer code 
\texttt{TLAC} \citep{Gronke_2014, Dijkstra_2014}, which is adopted for generating model \lya\ profiles. Finally, we introduce a novel machine-learning approach to effectively explore the multi-dimensional parameter space to infer the underlying physical quantities based on the observed line profiles.


\subsection{Spatially Varying \lya\ Profiles Observed in the Sunburst Galaxy}
\label{sec:quantifyinglyaspectraldifferences}

To assess the differences between the \lya\ spectra of three different images and to facilitate subsequent model comparisons, we first subtract the continuum in each spectrum using a best-fit linear model obtained from a $\chi^2$ minimization routine.  The best-fit continuum was determined using the observed spectrum within $\pm\,1400$ \kms, excluding the \lya\ emission feature at $\pm\,1000$ \kms.  In addition, individual \lya\ lines are normalized by the total integrated line flux within $\pm\,1400$ \kms\ after the continuum subtraction to enable direct comparisons of the underlying gas kinematics.  

The continuum-subtracted and line-flux normalized \lya\ profiles displayed in Figure \ref{fig:spectraoverview} show that all three images, 1.4, 1.10, and 200.8 display a prominent triple-peak \lya\ profile with a dominant central Gaussian component superimposed on a classic red-peak enhanced double-peak feature.  In addition, the \lya\ lines of images 1.4 and 1.10 show a remarkable agreement\footnote{A perfect correspondence between the spectra is nevertheless unlikely (even despite the high signal-to-noise and hence low Poisson noise of our spectra) since foreground \lya\ forest absorption can alter the observed profiles differently along the independent paths traced out by the light on its journey from the source to the observer.}, as demonstrated by the difference spectrum in the bottom panel of Figure \ref{fig:spectraoverview} and by the best-fit \lya\ radiative transfer models described in \S\ \ref{sec:lyaradiativetransfermodels} below. In contrast, the \lya\ line from image 200.8 shows a clear departure from the other two, both in terms of the shape of the double-peak profile and in terms of the relative strength of the central Gaussian component.  The observed differences provide clear empirical evidence in support of image 200.8 originating in a source separate from that of images 1.4 and 1.10 \citep[see also][]{Sharon2022}.

\subsection{\lya\ Radiative Transfer Models}
\label{sec:lyaradiativetransfermodels}

The resonant scattering nature of the \lya\ line, involving a random walk process, produces non-linear frequency shifts, which necessitates modeling the emergent \lya\ profile with simulations. Significant effort has gone into simulating \lya\ radiative transfer \citep[e.g.,][]{Dijkstra_2006, Verhamme_2015, Kakiichi_2021}. In this work, we utilize the radiative transfer code \texttt{TLAC} \citep{Gronke_2014, Dijkstra_2014} to simulate the diverse shapes of the \lya\ line. In dense \ion{H}{1} regions, \lya\ photons struggle to escape as they scatter repeatedly until they shift out of resonance. This results in a broad profile with minimal flux at the systemic velocity and widely separated blue and red peaks. As demonstrated by e.g. \cite{Verhamme_2006, Verhamme_2008, Verhamme_2015}, the \lya\ photons become frequency redistributed by the surrounding neutral hydrogen species and undergo a series of random scattering events. These scatterings cause the \lya\ profile to change depending on the underlying physical conditions of the surrounding gas, thus encoding the characteristics of the scattering medium encountered after leaving its birth site.

\begin{table*}\renewcommand{\arraystretch}{1.5}
	\centering
    \caption{Sampling grid for the \lya\ radiative transfer models used in this study.}
	\begin{tabular}{@{\hskip 0.25in}c@{\hskip 0.25in}|@{\hskip 0.5in}c@{\hskip 0.5in}}
    \hline
    \hline
    Gas Label & Grid Interval Configuration \\
    \hline
    \hline
    $\sigma_{\mathrm{int}} / (\mathrm{km} \, \mathrm{s}^{-1})$ & \makecell{Non-uniform sampling $\sigma_{\mathrm{int}} / (\mathrm{km} \, \mathrm{s}^{-1}) \, \in \, [0, 5, 10, 15, 20, 25, 30, 35, 40, 45, 50, 75, 100, 125,$ \\ $150, 175, 200, 225, 250, 275, 300, 325, 350, 375, 400, 425, 450, 475, 500]$} \\
    \hline
    $\mathrm{log}(N_{\mathrm{HI}}/\mathrm{cm}^{-2})$ & Uniform sampling $\mathrm{log}(N_{\mathrm{HI}}/\mathrm{cm}^{-2}) \, \in \, [18.0, 18.2, ..., 20.0]$ in steps of 0.2 dex\\
    \hline
    $\varv_{\mexp} / (\mathrm{km} \, \mathrm{s}^{-1})$ & Uniform sampling $\varv_{\mexp} / (\mathrm{km} \, \mathrm{s}^{-1}) \, \in \, [-300, -290, ..., 300]$ in steps of 10 \\
    \hline
    $\mathrm{log}\left(T/\mathrm{K}\right)$ & Uniform sampling $\mathrm{log}\left(T/\mathrm{K}\right) \, \in \, [3.6, 3.7, ..., 5.2]$ in steps of 0.1 dex \\
    \hline
    \hline
	\end{tabular}
     \label{tab:grid}
\end{table*}

\lya\ profiles consisting of more than a simple blue and red peak combination are common among galaxies with high LyC escape fractions \citep[see e.g.,][]{Naidu_2022}, and the Sunburst Arc is a prime example. In these cases, the observed \lya\ profiles likely result from a combination of two distinct escape modes \citep[e.g.,][]{Rivera-Thorsen2017}: (i) scattering, resonant escape through regions of high \ion{H}{1} column density, which produces red and blue peaks; and (ii) direct escape through porous channels with \lya\ opacity of $\tau_{\rm Ly\alpha} \lesssim 1$ and corresponding neutral hydrogen column density $N_{\rm HI}<1.7\times 10^{13}\,\cmjj$, which manifests as central \lyaper\ 

Motivated by previous work that modeled the triple-peaked morphology of \lya\ profile with an ionized channel in a thin spherical shell of neutral \ion{H}{1} gas around a central star-forming region sourcing the \lya\ \citep{Behrens_2014, Verhamme_2015, Rivera-Thorsen_2015, Rivera-Thorsen2017}, we simulate \lya\ profiles by initializing a dust-free thin spherical shell with a physical radius of $10\,\mathrm{pc}$ where the outer 10\% of the radius is filled with neutral hydrogen. This is commonly referred to as just the "shell model" \citep[see e.g.][]{Ahn_2004}. We vary the intrinsic dynamics of the \lya\ source, $\sigma_{\mathrm{int}}$, and the neutral hydrogen column density $N_{\mathrm{HI}}$, expansion/outflow velocity $\varv_{\mathrm{exp}}$, and effective\footnote{It is expected that both thermal and non-thermal motions contribute to the observed line-of-sight velocity dispersion.  Given a lack of constraints to isolate their respective contributions to the observed line widths, we consider a single effective temperature in the radiative transfer model.} temperature $T$ of the gas shell. We produce a grid of models for the parameter space outlined in Table \ref{tab:grid} for a total of $330{,}803$ simulated \lya\ spectral profiles. 

To summarize, each realization releases $10^5$ photons from the center of the shell whose velocities relative to the \lya\ rest frame are sampled from a Gaussian with a standard deviation set by $\sigma_{\mathrm{int}}$. This parameter encodes the intrinsic dynamics of the source. Each photon is emitted in a random direction from the central source and propagates through the simulation by scattering with the neutral hydrogen particles in the cloud (realized as cells) in a random walk until it escapes the cloud, in which it has undergone a measurable frequency shift. Generally, the photon escapes the cloud if its frequency has shifted out of resonance with the neutral hydrogen particles, or if the neutral hydrogen column density is low enough to significantly increase the mean free path of the \lya\ photons.

Following the convention, the sampled frequency shift for each photon, $x$, is related to its velocity shift, $\varv$, from the \lya\ rest frame in Doppler units according to
\begin{equation}
\label{eqn:freqshift}
    x = \frac{\nu - \nu_0}{\Delta \nu_{\mathrm{D}}} = - \frac{\varv}{b},
\end{equation}
where $\nu_0=2.47\times 10^{15}$ Hz is the \lya\ resonant frequency, $\Delta \nu_{\mathrm{D}} = (\varv_{\mathrm{th}}/c) \nu_0$, and the Doppler parameter is given by the bulk flows, $\varv_{\mathrm{turb}}$, and the thermal velocity, $\varv_{\mathrm{th}}$, following
\begin{equation}
\label{eqn:Doppler}
    b = \sqrt{\varv^2_{\mathrm{th}} + \varv^2_{\mathrm{turb}}}.
\end{equation}
We then convert the frequency shift of each photon to its velocity in the \lya\ rest frame defined by the redshift of the Sunburst Arc following

\begin{equation}
\label{eqn:voffset}
    \begin{split}
        \varv   & = - \left( \frac{2 \, k_{\mathrm{B}}}{m_{\mathrm{H}}} \right)^{1/2} \, \cdot \, x \, \cdot \, T^{1/2} \\
            & \approx -12.8 \, \cdot \, x \, \cdot \, (T/10^4\,{\rm K})^{1/2} \, \mathrm{km} \, \mathrm{s}^{-1},
    \end{split}
\end{equation}
where $T$ is the effective temperature of the \texttt{TLAC} model.  We adopt $z=2.37094$ from \cite{Rivera-Thorsen2017} as the systemic redshift of the Sunburst galaxy.
All photons in each run are then binned by their velocities in bins of $2 \, \mathrm{km} \, \mathrm{s}^{-1}$ in the range $[-1400, 1400] \, \mathrm{km} \, \mathrm{s}^{-1}$.

\subsection{Constraints on Model Parameters and Associated Uncertainties}
\label{sec:constraintsanduncertainties}

Our analysis seeks to hone in on a physical picture of the conditions of the gas that allow \lya\ photons to escape from the parent star-forming regions in Sunburst while keeping the number of required spectral features as few and simple as possible. This is built on the approach from previous work that some \lya\ flux undergoes significant scattering on its path through an optically thick gaseous medium before escaping in the outskirts of the \ion{H}{1} cloud \citep[see e.g.][for a review]{Dijkstra_2017}. For nearby galaxies, clear evidence for this resonant scattering has been observed and is commonly known as \lya\ halos \citep[e.g.,][]{Ostlin_2014}. While significant \lya\ flux is expected to undergo frequency shifts from scattering processes in high-density gas, some \lya\ flux may originate from ionized channels/chimneys in the gas enabling direct escape without frequency redistribution. 

Our analysis aims to employ resonant scattering models over a continuous range of physical parameters that can be randomly sampled using a Monte Carlo Markov Chain (MCMC).  In addition, we incorporate an additional component to allow for direct escape of \lya\ near the rest-frame velocity. In the case of the Sunburst Arc, we wish to understand the origin of the triple-peaked \lya\ feature present in both our LyC leaking and non-leaking regions. While triple peaks can be simulated through multiple incidences of backscattering \citep[see e.g.,][]{Verhamme_2006}, such backscattering is expected to impose a velocity offset from the source and produces an asymmetric peak. It therefore would not explain the observed central peak in Figure \ref{fig:spectraoverview}.

To match the triple-peaked shape of the \lya\ profiles in our sample, we assume a two-component fitting procedure: a scattered \lya\ component and a single Gaussian component centered around the \lya\ rest frame, designated "the central Gaussian". To allow for sufficient leeway in the fitting of these components, we treat the velocity centroids of the scattered component and the central Gaussian as free parameters, $\varv_{\mathrm{Ly}\alpha}$ and $\varv_g$, respectively. These parameters help quantify the relative kinematics between different sources of \lya\ photons in Sunburst.

To assess the uncertainties of the best-fit model, we construct posteriors of individual model parameters using 
MCMC. This requires access to a continuous high-dimensional parameter space spanned by the free parameters necessary to characterize the \lya\ resonant scattering component of the profile. However, simulating radiative transfer for $10^5$ photons in a single trial is already computationally intensive, especially at higher \ion{H}{1} column densities. The need to increase the efficiency of uncertainty estimates over a vast high-dimensional parameter space motivated us to develop a novel machine-learning method to interpolate the parameter space between discrete model grid points summarized in Table \ref{tab:grid}. We first train a neural network to learn the \lya\ resonant scattering and to accurately predict the shape of the emergent profile based on a set of physical parameters. The training process for the neural network to emulate \lya\ radiative transfer spectra from simulations is further explained in Section \ref{sec:neuralnetwork}.

We adopt a Bayesian framework for our uncertainty estimates and we choose broad flat priors around our best-estimated initial input. For our set of free parameters, $\boldsymbol{\Theta}=(\sigma_{\mathrm{int}}, N_{\mathrm{HI}}, \varv_{\mathrm{exp}}, T, \varv_g, \sigma_g, \varv_{\mlya})$, the likelihood function, $\mathscr{L}$, is defined by
\begin{equation}
\label{eqn:loglikelihood}
\begin{aligned}
    &\mathscr{L} (\boldsymbol{\Theta}) \\ & = \prod_{i} \exp \Bigg\{ - \frac{\mathrm{m}(\lambda_i) \, [D(\lambda_i) - M(\lambda_i | \boldsymbol{\Theta})]^2}{2 S^2(\lambda_i)} 
    \Bigg\},
\end{aligned}
\end{equation}
where $D(\lambda_i)$ and $M(\lambda_i)$ are, respectively, our observed and model flux values at wavelength $\lambda_i$ of pixel $i$, $S(\lambda_i)$ is the corresponding error flux, and $\mathrm{m}(\lambda_i)$ is a mask with values of either 0 or 1.
The mask $\mathrm{m}(\lambda_i)$ is applied to minimize the influence of small-scale features likely due to foreground \lya\ forest or absorption by material falling in or out of the galaxy (see \S\ \ref{sec:foregroundabsorbers}). This is necessary particularly because of the high-quality and high-resolution MIKE spectra with extremely high $S/N$, revealing small-scale features that were not previously seen in low-resolution spectra, such as those from MagE with $R \approx 4700$ \citep[][]{Rivera-Thorsen2017, Mainali_2022, Owens_2024}.  The high $S/N$ data
 led to unrealistically small statistical uncertainties for the model parameters. For example, the MCMC returns uncertainties of $\sim 0.05 \, \mathrm{dex}$ for $\mathrm{log}(N_{\mathrm{HI}}/\mathrm{cm}^{-2})$ and $\mathrm{log}(T/\mathrm{K})$, and $\lesssim 1$ for $\varv_g / (\mathrm{km} \, \mathrm{s}^{-1})$ and $\varv_{\mlya} / (\mathrm{km} \, \mathrm{s}^{-1})$. By iterating with adding one mask at a time, we can quantify the significance of systematic uncertainties in the best-fit model parameters. The masked velocity windows are listed in Table \ref{tab:mask_ranges}. Note that for images originating in the same source, namely images 1.4 and 1.10, the same set of masks are applied. Each mask is added progressively from top to bottom with each new iteration, so the final iteration incorporates all the masks applied in earlier steps.

\begin{table}
\renewcommand{\arraystretch}{1.5}
\centering
\caption{Mask velocity windows in \lya\ profiles.}
\label{tab:mask_ranges}
\begin{tabular}{c|c|c}
\hline
\hline
Mask ID & \multicolumn{2}{c}{Range [$\mathrm{km} \, \mathrm{s}^{-1}$]} \\
\hline
 & 1.4 and 1.10 & 200.8 \\
\hline
\hline
\#1 & [$-1000$, $-230$] & [$-1000$, $-200$] \\
\#2 & [$400$, $460$] & [$385$, $455$] \\
\#3 & [$110$, $145$] & [$110$, $145$] \\
\#4 & [$35$, $40$] & [$40$, $48$] \\
\#5 & [$40$, $47$] & [$35$, $40$] \\
\#6 & [$30$, $35$] & [$48$, $55$] \\
\#7 & [$47$, $55$] & [$30$, $35$] \\
\#8 & [$22$, $30$] & [$55$, $60$] \\
\#9 & [$55$, $60$] &  \\
\hline
\hline
\end{tabular}
\end{table}

Finally, the two-component model, incorporating a central Gaussian and a scattered \lya\ feature, is normalized by the total integrated flux within the velocity window of $\pm\,1400$ \kms, and the relative contributions of individual components are scaled by their normalization coefficients determined based on a $\chi^2$ minimization. These coefficients thus correspond to the relative strengths of the central Gaussian and scattered \lya\ components.

\subsection{A Neural Network to Emulate \lya\ Radiative Transfer Simulations}
\label{sec:neuralnetwork}

For this work, we adopt a machine learning approach to facilitate the uncertainty estimates of the best-fit \lya\ model parameters based on an MCMC routine.  Neural networks have previously been used to interpolate between the parameter grids of \lya\ profiles, as demonstrated by the \texttt{zELDA} software \citep{Gurung-Lopez:2022}. While \texttt{zELDA} incorporates dust and assumes a fixed effective temperature, our approach allows the effective temperature to be a free parameter but does not include dust effects. This enables our neural network to cover a range of physical conditions dictated by the temperature and turbulence of the gas, albeit without considering dust-related factors.

Our neural network code \texttt{CCLya-Payne}\footnote{Available for download at \href{https://github.com/highzclouds/CCLya-Payne}{https://github.com/highzclouds/CCLya-Payne}.} is designed to fit physical parameters of a shell model from observed \lya\ profiles. \texttt{CCLya-Payne} enables continuous access across the parameter space outlined in Table \ref{tab:grid}, making it well-suited for MCMC fitting. This code is a modified version of the neural network architecture of \texttt{ThePayne} \citep{Ting2019}. While \texttt{ThePayne} is designed to determine properties of stellar atmospheres by fitting absorption lines, we adapt the code to fit \lya\ emission profiles. This requires modifying the code from fitting stellar atmospheric parameters (elemental abundances, radial velocity, etc.) to emulate \lya\ profiles from \texttt{TLAC} for our model grid. Our final setup is a simple fully connected neural network with three hidden layers.

From the $330{,}803$ simulated \lya\ spectra, covering the parameter space outlined in Table \ref{tab:grid}, we randomly select 75\% ($248{,}102$) as training spectra and reserve the remaining 25\% ($82{,}701$) as validation spectra for the neural network training. The validation set serves as an independent assessment of the neural network's performance in emulating the spectra.
Physically, our emission spectra should be positive for all $\varv$-values. To ensure only positive fluxes in the output, we apply a sigmoid activation function,
\begin{equation}
\label{eqn:sigmoid}
    \sigma_{S} (f) = \frac{1}{1 + e^{-f}}
\end{equation} to the input flux, $f$, before the output layer. 

The neural network is trained by feeding it normalized input spectra, so scaling the sigmoid by an arbitrary scale factor is not necessary. Activation functions serve to introduce sufficient non-linearity to the neural network, and we leave the first two layers of the network as LeakyReLU functions,
\begin{equation}
\label{eqn:leakyrelu}
    \sigma_{LR} (f) = \begin{cases}
    0.1\,f, \; \; f < 0\\
    f, \; \; \geq 0\\
    \end{cases}
\end{equation}

The LeakyReLU function in Equation \ref{eqn:leakyrelu} is preferred over a traditional ReLU function because of its non-zero gradient below $x < 0$ which would otherwise limit the training speed of the neural network.

Our final neural network architecture is described by Equation \ref{eqn:nn_function}, which given a fully trained set of weights, $w$, and biases, $B$, emulates the flux per velocity bin, $f_{\varv}$, as a function of the input parameters, $\ell$, of a \lya\ resonantly scattered profile:

\begin{equation}
\label{eqn:nn_function}
    f_{\varv}(\ell) = w_4 \sigma_{S}\{ w_3\sigma_{LR}[ w_2 \sigma_{LR} ( w_1 \, \ell + B_1) + B_2 ] + B_3 \} + B_4
\end{equation}
$\sigma_{S}$ and $\sigma_{LR}$ are the sigmoid and LeakyReLU activation functions defined in Equations \ref{eqn:sigmoid} and \ref{eqn:leakyrelu}, respectively. This enables the model to accurately capture the intricate characteristics inherent in the \lya\ profile, ensuring a robust emulation of the profile's response to parameter variations. One can in principle increase the number of neurons, $k$, to increase the complexity of the neural network. However, increasing the number of neurons and activation functions is prone to overfitting. We set $k = 300$ for all four layers.

During the training process, the neural network minimizes the loss function $L^1$ \citep[see e.g.,][]{Tibshirani:1996} for the training spectra. This loss function is a mean absolute error (MAE) calculation defined as
\begin{equation}
\label{eqn:l1}
    L^1 = \frac{1}{N} \, \sum_{i=0}^N |f_i - \Bar{f_i}|
\end{equation}
where $f_i$ is the "true" 
spectral value and $\Bar{f_i}$ is the spectral value emulated by the neural network at pixel $i$. The loss is summed over the total number of pixels, $N$.

\begin{figure}
	\includegraphics[width=0.9\columnwidth]{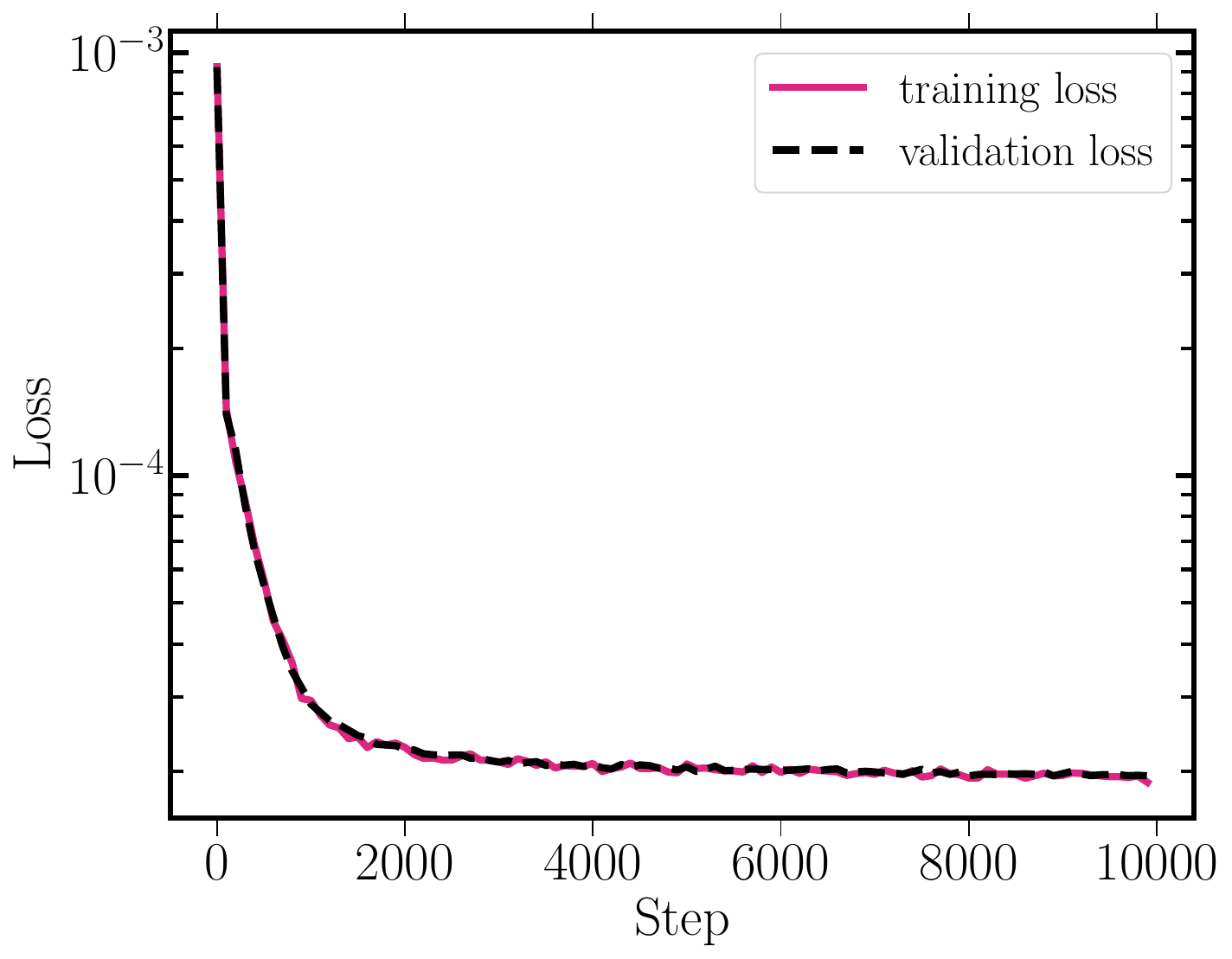}
    \caption{Training and validation loss for our neural network as a function of training steps. By comparing the two losses, we observe that they decrease consistently. This suggests that the neural network avoids overfitting the training spectra.}
    \label{fig:loss}
\end{figure}

The loss calculated for the validation spectra independently evaluates how well the neural network emulates the simulation of the spectra. If the neural network overfits the spectral variation, the loss of the validation spectra will be worse relative to the loss of the training spectra. Then, to avoid overfitting, we require the validation loss to decrease as a function of the neural network training steps at the same rate as the training loss. We confirm that our neural network avoids overfitting in Figure~\ref{fig:loss} by tracking the loss function (MAE\footnote{While Figure \ref{fig:loss} shows the result based on an MAE calculation, we have also experimented with adopting a mean square error (MSE) loss function $L^2$ and found that the neural network shows a consistent loss evaluation across the parameter spaces.}, defined in Equation \ref{eqn:l1}) for both the training and validation set separately. The results of our \lya\ spectral analysis are presented in \S\ \ref{sec:results}.

\section{Gravitational Lens Model}
\label{sec:gravitationallensmodel}

A model of the foreground cluster mass distribution is necessary to reconstruct and interpret the morphology of the host galaxy in the source plane. Models of this kind are generally referred to as gravitational lens models. All lens models come with assumptions of the invisible mass distribution and perturbers within it, and all models can be treated differently to optimize for global or local image constraints. For the Sunburst Arc, several previous works have attempted to model the foreground mass distribution using multiple images of discrete sources in the host galaxy (see Figure \ref{fig:lens} for a summary of these images and Table \ref{tab:nomenclature} for references of previous studies).  However, a caveat is a lack of agreement in the origins of image 200.8 (referred to as "Godzilla" by \citealt{Diego_2022}), and images 300.80 and 300.81  (referred to as the "Pair" by \citealt{Diego_2022}). While the "Godzilla" and the "Pair" were treated as originating in separate sources by \cite{Diego_2022}, all three images were treated as originating in one source (Source 4 in the discussion below) by \cite{Sharon2022}.
Our motivation for making an independent lens model is to experiment with allowing image 200.8 to arise in a different star-forming region from the source responsible for the "Pair" at $0.4''$ away\footnote{Obtaining an independent confirmation of the origins of the "Godzilla" and the "Pair" based on spectroscopy is challenging. A primary caveat is that both images of the "Pair" are relatively faint with $AB({\rm F606W})\!\approx\!23.1$, while the "Godzilla" is brighter with $AB({\rm F606W})\!\approx\!22.1$. Treating these as separate sources is at least justified by their different colors with $AB({\rm F814W})-AB({\rm F160W})\!\approx\!-0.4$ and $0.0$ for the "Godzilla" and the "Pair", respectively \citep[e.g.,][]{Sharon2022}. We performed this photometry using a $0.15"$ aperture set by the size of "Godzilla" in the filter with the worst seeing, F160W, and applying it to "Godzilla" and the "Pair" separately in each filter.}. 
The goal is to reconstruct 
the overall spatial distribution of individual star-forming regions in the source galaxy.

We additionally choose to only include image constraints with spectroscopically confirmed redshifts. We define a reference coordinate frame using the Brightest Cluster Galaxy (BCG), and  
the coordinates of all objects are defined in terms of their $\Delta \mathrm{RA}, \Delta \mathrm{DEC}$ relative to the BCG at $\mathrm{RA}_{\mathrm{BCG}}, \, \mathrm{DEC}_{\mathrm{BCG}}$. Image constraints for our model and equations for $\Delta \mathrm{RA}, \Delta \mathrm{DEC}$ are presented in Appendix \ref{appx:image_constraints}.
We employ \texttt{GLAFIC} \citep{Oguri_2010} to model the mass distributions in three planes between the observer and the source (the Sunburst) at $z=0.4430$, 0.5578, and 0.7346.  

Our 
lens model is optimized for the areas of the NW arc covered by two of our three spectrograph pointings, namely images 1.10 and 200.8. Since two of the foreground galaxies are positioned close to the NW arc, their gravitational potential can lead to noticeable distortions to the images along the NW arc. Therefore, we consider these two galaxies as components in separate lens planes, positioned at $z=0.5578$ and $z=0.7346$ based on their spectroscopic redshifts. These halos have been approximated by previous models as galaxies in the same lens plane as the rest of the foreground cluster galaxies at $z=0.4430$ \citep[e.g][]{Sharon2022}. This arises from the challenge of the observed position of galaxies in background lens planes being warped by the foreground lens planes. Not all softwares have the ability to account for multiple lens planes. However, \texttt{GLAFIC} allows for calculating mass distributions in multiple lens planes, so we adopt the use of three lens planes in our gravitational lens model. We later describe our strategy to account for these background galaxies' mass distributions while still recovering their observed positions. Our model is the first to implement multiple lens planes for the Sunburst Arc. Appendix \ref{appx:lensmodelparameters} gives an overview of each lens model component and their best-fit parameters.

\subsection{Primary Lens Plane Components}
\label{sec:primarylensplane}

The majority of the foreground mass distribution is located in the plane at the redshift of the cluster PSZ1 G311.65-18.48, $z = 0.4430$, hereby referred to as the primary lens plane. 
In this subsection, we describe the model construction of the primary lens at $z = 0.4430$.

\subsubsection{Individually Optimized Cluster Halos}
\label{sec:individualhalos}

\texttt{GLAFIC} is a parametric lens modeling code that constructs the total mass distribution of the foreground galaxy cluster based on a combined mass distribution profile of multiple halos. In our model, we adopt the pseudo-jaffe ellipsoid specified by the \texttt{jaffe} parameter in \texttt{GLAFIC} 
for the BCG and the mass components of all photometrically identified cluster members. We use broad priors for each free parameter in the halo.
Based on the location of the arcs, we adopt a global dark matter (DM) halo which we allow varying within a broad region just northeast of the BCG, referred to as the "DM Halo". As is common practice in lens modeling, we assign a smaller halo to the BCG itself, the "BCG Subhalo", which can vary within a box of $2''\!\times\!2''$ around the BCG centroid. To accurately reproduce images 1.4, 1.5, and 1.6 in the eastern part of the N arc, we add a subhalo east of the N arc which we name the "NE Subhalo". Motivated by the need to remove ghost images (image predictions that appear without observed counterparts) north of the SE arc, we introduce a halo at roughly $11''$ E and $40''$ S (corresponding to an angular distance of $41''$ from the BCG center) and we refer to this as the "S Subhalo". We set the initial position for this halo to the location of a bright cluster galaxy in this area and let it vary within a box of $10''\!\times\!10''$ centered at this location.

\begin{figure*}
\centering
 \makebox[\textwidth]{\includegraphics[width=0.75\paperwidth]{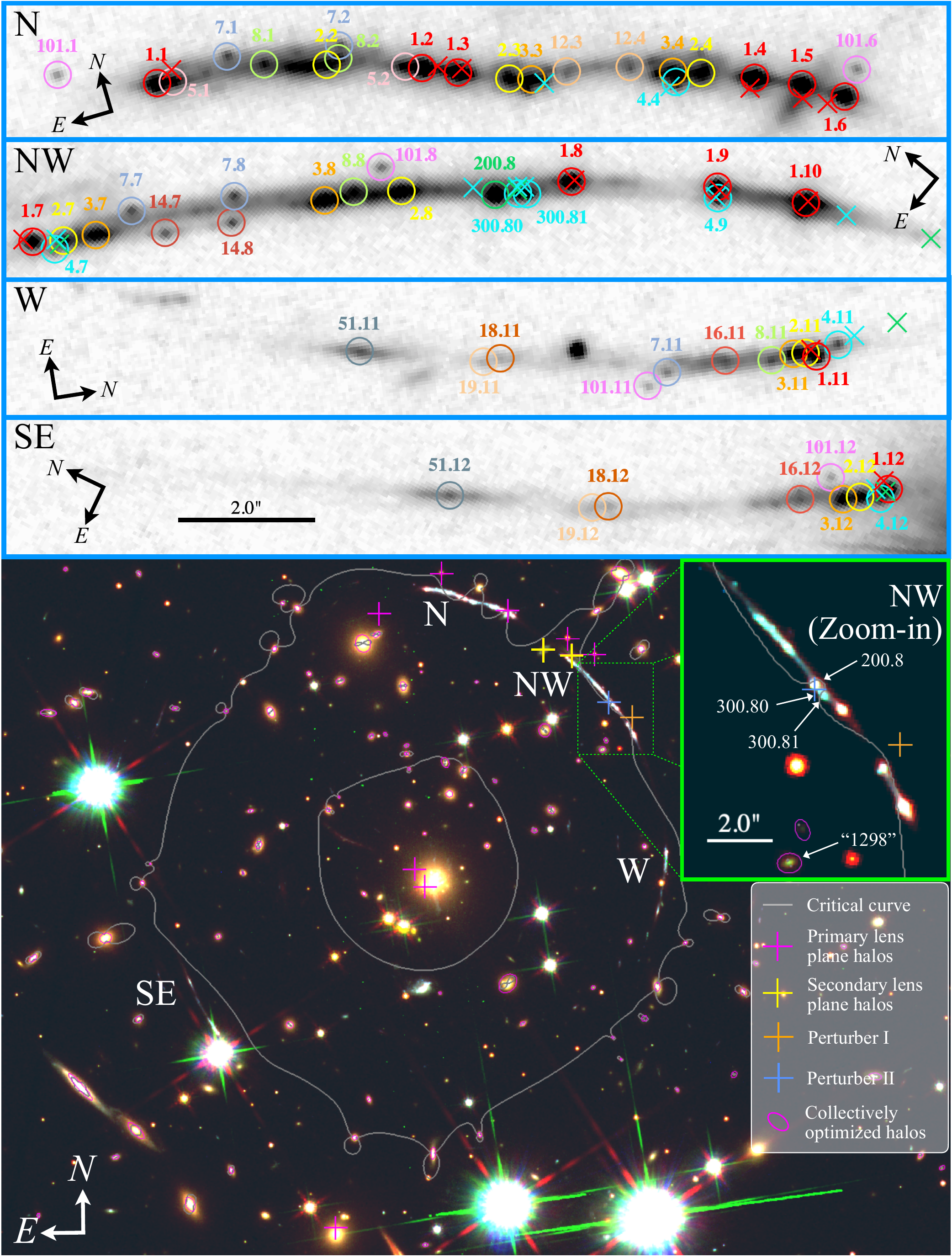}}
    \caption{\textit{Top} four panels: {\it HST} F606W images of individual arcs, illustrating the multiple images included in our lens model. Circles show the observed locations of each image 
    and crosses show the relensed locations of Source 1 (red), 4 (cyan), and 200 (green). \textit{Bottom:} Composite color image, comprising \textit{HST} F160W, F814W, and F606W images, of the cluster field with the best-fit critical curve of sources at $z\!=\!2.37$ shown in gray line. Magenta ellipses outline foreground cluster galaxies included in the \texttt{GLAFIC} \texttt{gals} component; magenta crosses indicate independently optimized halos; blue and orange crosses indicate the locations of Perturber I and II, respectively, and yellow crosses indicate secondary lens galaxies at $z\!\approx\!0.56$ and 0.73. 
    The inset shows a zoom-in region of images 8, 9, and 10 to display the effect of two dark perturbers in the NW arc. 
    }
    \label{fig:lens}
\end{figure*}

In order to produce the high image multiplicity of Source 1 in the N arc, we add a halo which we allow to vary within a box of $3''\!\times\!3''$ around the spiral galaxy near image 1.5, referred to as the "Spiral on N Arc". We allow the halo of an elliptical galaxy north of the N arc, which we call the "Elliptical N of N Arc", to vary independently from the galaxies in \texttt{gals}, but we fix its position to its measured centroid. Additionally, we free two elliptical galaxy halos north of the NW arc while fixing their positions to allow the critical curve to run between the mirror images of Source 14. These are referred to as "E Elliptical N of NW Arc" and "W Elliptical N of NW Arc".  The individually optimized halos above are marked by a magenta cross in Figure \ref{fig:lens}.

In addition, two small dark matter perturbers with no galaxy counterparts identified are necessary to explain the multiple images observed along the N and NW arcs.  We adopt the singular isothermal ellipsoid specified by the \texttt{sie} parameter in \texttt{GLAFIC}. The mass profiles of these halos are individually optimized.  These perturbers are marked by an orange and blue cross, respectively, in Figure \ref{fig:lens} (bottom panel).  
For collectively optimized foreground cluster members, we utilize the \texttt{gals} variable which is further explained in Section \ref{sec:gals}.

\subsubsection{Collectively Optimized Member Galaxies of the Cluster}
\label{sec:gals}

The individual masses of galaxy cluster members play a crucial role in shaping the gravitational potential of the lensing cluster. However, individually optimizing the mass distribution of every cluster member is computationally expensive. When their relative brightness can be determined, they can instead be modeled effectively by scaling their masses according to their measured visible light. Therefore, we perform source detection and photometry of cluster members with \texttt{Source Extractor} \citep{Bertin_1996} in dual-image mode to create a photometric catalog of potential cluster members in the F814W and F606W filters. We choose these bands because they encompass the $4000$ \AA\ break characteristic of passive galaxies at the cluster's redshift. We use F814W as the detection image and apply the apertures identified in the detection image to extract \texttt{MAG\_ISO} from F814W and F606W. We apply a detection threshold of $1.7\sigma$ relative to the background and set the minimum deblending contrast parameter to 0.005. The image frames are smoothed by a narrow $5\times5$ kernel of $\mathrm{FWHM} = 3.0$ pixels to avoid spurious detections.  

We remove foreground Milky Way stars by sorting by the objects' stellarity which is determined by a neural network algorithm in \texttt{Source Extractor}. All objects with $\texttt{CLASS\_STAR}\!>\!0.98$ are removed. We perform a selection of candidate cluster members following the Red Sequence method \citep[e.g.,][]{Gladders_2000} making color cuts with objects with $0.75 < AB(\mathrm{F606W}) - AB(\mathrm{F814W}) < 1.5$ and magnitude cuts for objects with $ 18 < AB(\mathrm{F814W}) < 24$. Finally, we visually review the galaxy catalog to eliminate remaining non-cluster members, e.g., stars. Our final sample of cluster members comprises $205$ galaxies.

We calculate the $\Delta \mathrm{RA}$ and $\Delta \mathrm{DEC}$ of each object in the catalog relative to the BCG coordinates. We also input the ellipticity, 
position angle (which is defined East of North), and relative luminosity based on the observed F814W magnitude 
into the catalog to inform the \texttt{gals} parameter in \texttt{GLAFIC}. 
Each collectively optimized galaxy is modeled by a pseudo-Jaffe ellipsoid profile.  
These collectively optimized member galaxy halos are marked by magenta ellipses in Figure \ref{fig:lens} (bottom panel) set by their semi-major and semi-minor axes and position angle calculated with \texttt{Source Extractor}.

\subsubsection{NW Arc "Perturber" Halos}
\label{sec:perturberhalos}

The NW arc has the most complex mass distribution of the four large arcs of the Sunburst Arc due to the presence of two foreground galaxies behind the lensing cluster 
and the lack of visible foreground cluster galaxies to explain the high image multiplicity of Source 1 near the western end of the arc (see Figure \ref{fig:lens}). We have attempted to create an accurate reproduction of the foreground mass in the northern part of the NW arc by modeling galaxies in multiple lens planes (see Section {\ref{sec:secondarylensplane}). However, the southern part of the NW arc is particularly challenging as the lack of visible foreground galaxies in its vicinity makes it challenging to wind the critical curve around these images, which is necessary to reproduce the correct image multiplicity. 

A couple of workarounds for modeling this particular part of the Sunburst Arc have been suggested. One such procedure was performed by \cite{Pignataro_2021} by individually optimizing the halo of a galaxy south of the NW arc (dubbed "1298"; see the inset of Figure \ref{fig:lens}) to allow the critical curve to go around image 1.9. Another method, employed by \cite{Diego_2022}, is to constrain the critical curve to pass through points at which the parity of the multiple images is expected to change. \cite{Diego_2022} introduced a small dark matter halo, which we will refer to as a "perturber", near image 200.8.  The inclusion of a perturber is found to reproduce the extreme magnification of 200.8 and the multiplicity of a single source giving rise to images 300.80 and 300.81. 

Motivated by the success of \cite{Diego_2022}, we choose a similar strategy for winding the curve around image 1.9 and matching the double images 300.80 and 300.81 as a result of the multiplicity of a single source through lensing magnification (see the inset of Figure \ref{fig:lens}). We adopt a parametric approach for the dark matter perturber by modeling it as a singular isothermal ellipsoid near image 1.9 and between 200.8, and the pair of 300.80 and 300.81. We pass the initial guess position for Perturber I at the position of image 1.9, and we let its position vary in a $1.0''\!\times\!1.0''$ box centered on this position. This Perturber I can be considered an extension of the halo of galaxy "1298" in \cite{Pignataro_2021} and reflects the uncertainty in the DM distribution in this region of the NW arc (see Figure \ref{fig:lens}). 

For the position of Perturber II, we calculate the midpoint between 200.8 and the midpoint between 300.80 and 300.81.
We extend a line between this point and the BCG reference coordinate and place our Perturber II $0.15"$ southeast along this line. This is to ensure the critical curve of this mass distribution (which we set to have zero ellipticity) passes through 200.8 and the midpoint between 300.80 and 300.81 (blue cross in Figure \ref{fig:lens}).  Incidentally, recent deep images obtained from the {\it James Webb Space Telescope} \citep{Gardner_2006} have revealed a low surface brightness galaxy under image 200.8 that may be associated with Perturber II \citep[see][for details]{Choe:2024}.

To optimize our lens model for the NW arc, where the perturber components play a significant role in shaping the critical curve and determining the multiplicity of images, we perform a second iteration on our lens model. After the initial optimization of all halos, we fix the parameters of every halo except Perturber I and II, which are allowed to vary. This approach allows us to refine the influence of these perturbers independently, ensuring the most accurate model for the mass distribution around the NW arc. The best-fit parameters derived from this iterative process constitute our fiducial lens model, as listed in Table \ref{tab:lens_parameters}.

\subsection{Secondary Lens Plane Components}
\label{sec:secondarylensplane}

We determine the precise angular positions of the spirals identified at $z=0.5578$ and $z=0.7346$ by ray tracing their observed positions to their respective redshifts. Given the complexity of the lensing scenario involving two secondary lens planes (with the lens plane at $z=0.4430$ conventionally termed the "primary" lens plane), we undertake two iterations of ray tracing:
\begin{enumerate}
    \item Initially, we employ the deflection field from the single primary lens plane at $z=0.4430$ to ray trace the source location of the spiral at $z=0.5578$.
    \item Subsequently, we utilize the deflection field from both the primary lens plane at $z=0.4430$ and the secondary lens plane at $z=0.5578$ to pinpoint the position of the spiral at $z=0.7346$.
\end{enumerate}
The spiral centroids, as predicted by the \texttt{lenscenter} command in \texttt{GLAFIC}, align precisely with the observed locations of the spirals in the \textit{HST} images. We represent this alignment visually with yellow crosses in the NW arc in Figure \ref{fig:lens}. The two halos located in the secondary lens planes are parametrized as \texttt{jaffe} profiles, as we do for the majority of the primary lens plane components.

\subsection{Source-Plane Reconstruction}
\label{sec:sourceplanereconstruction}

The best-fit gravitational lensing model, involving multiple lens planes, allows us to reconstruct the image of Sunburst in the source plane. The main goal is to obtain quantitative constraints for the relative separations between individual star-forming regions in Sunburst. 
We adopt Source 1, the LyC leaker, as the reference point, 
because Source 1 is the only source that appears in all images.  In addition, the anchor position between different images of Source 1 is set to the delensed location from image 1.10. 

\begin{figure}
\centering
	\includegraphics[width=0.925\columnwidth]{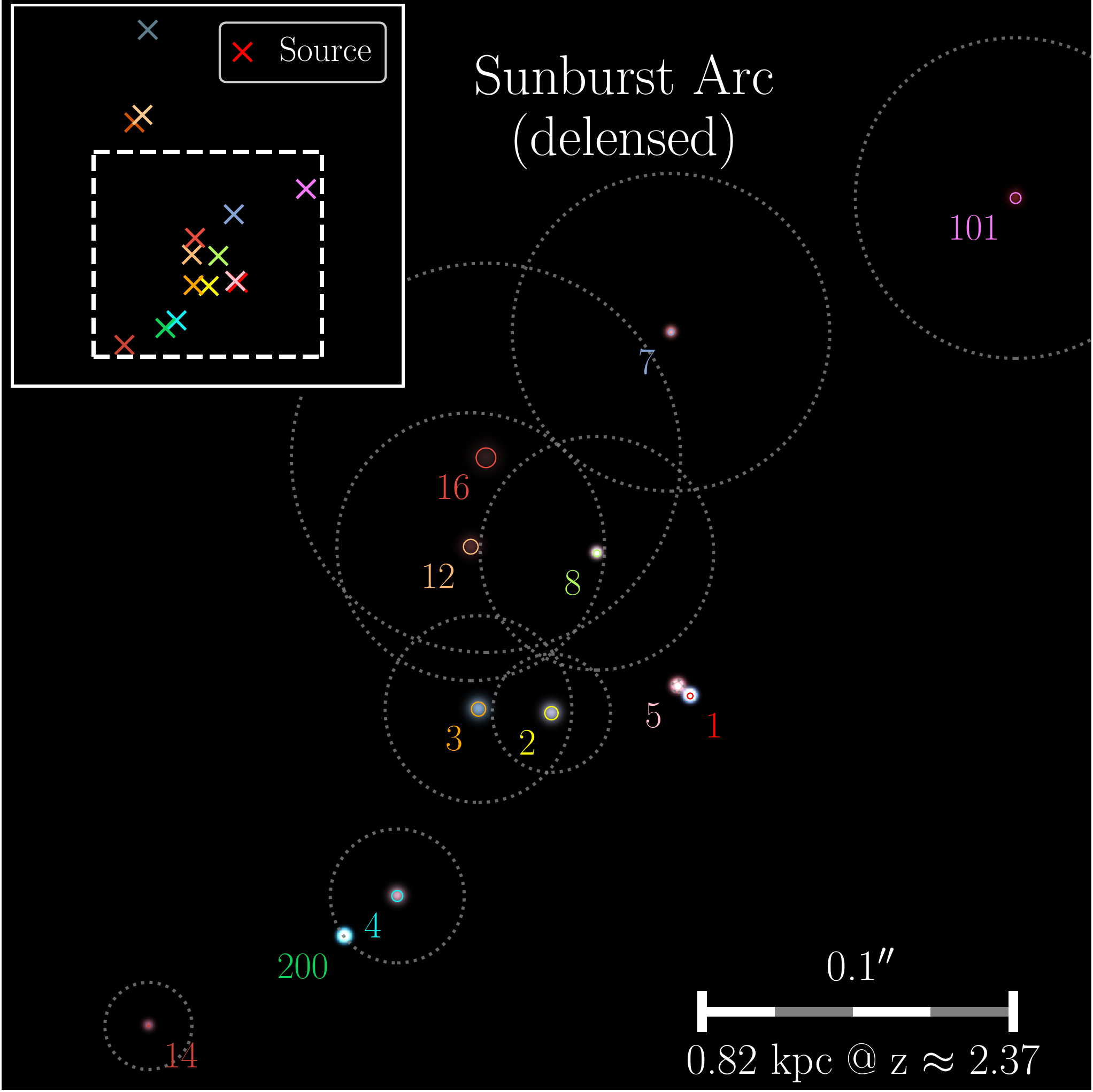}
    \caption{Composite color image of individual star-forming regions reconstructed in the source plane of Sunburst at $z \approx 2.37$.  The image is formed by combining delensed \textit{HST} images in the F606W, F814W, and F160W filters based on our fiducial lens model. The inset in the upper-left corner displays all gravitationally magnified regions, including three more sources outside the main panel. We anchor the position of each source relative to Source 1, the brightest in Sunburst and a LyC leaker. The labels indicate the IDs of individual sources along with 
    a dotted gray circle representing the scatter in the source position inferred from multiple images.}
    \label{fig:sourceplane}
\end{figure}

The computation of the source-plane coordinates of an image is commonly referred to as \textit{delensing}. We have attempted to ray trace the corners of each pixel back to the source plane and reconstruct each pixel in the source plane. However, in regions proximal to the critical curve the high magnification elongates a delensed square pixel from the image plane into a narrow diamond shape in the source plane. This elongation arises from the substantial difference in magnificaiton along the tangential and radial axes, and this is a fundamental limitation of the point-spread-function (PSF) of the image along the radial direction. As a result, the delensed images do not retain the maximal spatial information of each source along the radial direction of the lens.

To construct a source-plane image of the Sunburst galaxy for quantifying the physical separations between individual star-forming regions, we adopt an aperture size based on a minimal demagnified effective radius determined along an arc for each lensed image in the source plane and place the observed flux within the effective radius in the aperture after correcting for the aperture difference to conserve the surface brightness.  For sources with multiple images, we repeat this exercise for all lensed images and determine a final source-plane position and flux, as well as uncertainties, based on the mean and dispersion over all images.

To produce a source-plane composite color image, we repeat the delensing exercise for all three bands, including F606W, F814W, and F160W. Individual source positions in F814W and F160W are registered to their positions observed in F606W.  
The final reconstructed source-plane image of the Sunburst galaxy is presented in Figure \ref{fig:sourceplane}, summarizing the relative locations of individual star-forming knots and associated uncertainties relative to Source 1.  The RMS scatter of each delensed image is indicated by a gray dotted circle.  The point with no circle, Source 200, has only one image observed, namely image 200.8 (the "Godzilla" from \citealt{Diego_2022}).  The distances and uncertainties 
of individual sources to Source 1 are listed in Table \ref{tab:separations}.
The relative orientations of the delensed sources are all visually consistent with the predicted source plane in \cite{Sharon2022}. Additionally, the separations between Source 1 and most other sources (see Table \ref{tab:separations}) agree within the error bars reported by \cite[][see their Table 4]{Sharon2022}, except for Source 7, 16, 18, 19, and 51. Notably, our model treats image 200.8 ("Godzilla") as a separate source rather than being associated with Source 4, as done in \cite{Sharon2022}. This distinction informs our interpretation of the spatial offset between the LyC leaker and non-leaker in this work, and we confirm that this treatment does not significantly alter the predicted source-plane separation between Source 1 and "Godzilla" compared to \cite{Sharon2022}, as we find the same separation within the uncertainties (see \S\ \ref{sec:results}).

By ray-tracing the source-plane position of Source 4 to the image plane, our gravitational lens model predicts the existence of three distinct images of Source 4, each within 1 arcsecond of 200.8, 300.80, and 300.81. This implies that the "Pair" (300.80 and 300.81) may be multiple images of Source 4. Additionally, the third image predicted in this area suggests that Source 200 may also be a multiple image of the same source. In the subsequent analysis, we consider images 300.80 and 300.81 as two of the multiple images of Source 4 but leave image 200.8 as an independent source given its extreme brightness and optical color, distinct from Source 4 seen in {\it HST} images (see \S\ \ref{sec:gravitationallensmodel} and also \citealt{Sharon2022}).  

Figure \ref{fig:sourceplane} illustrates the exceptional spatial resolving power afforded by strong gravitational lensing of a distant starburst galaxy with a spatial resolution of $\lesssim 0.02''$, corresponding to a physical scale of $\lesssim 160$ pc at $z\approx 2.4$.  It helps reveal a clumpy medium with spatially confined LyC leakage on scales less than 100 pc \citep[see also][]{Sharon2022}.

\begin{table}
\renewcommand{\arraystretch}{1.5}
\centering
\caption{Summary of Sunburst source-plane reconstruction.}
\label{tab:separations}
\begin{tabular}{c|c|c|c}
\hline
\hline
Source ID & \# of Images & Predicted Distance [kpc] \footnote{Mean distance to Source 1 and the associated standard deviation computed based on the reconstructed source-plane positions of all available images.  For sources with only two images found, the distance measurements from individually reconstructed source positions are listed.} & RMS Scatter [kpc] \footnote{RMS scatter of the source positions inferred from multiple delensed source positions.  For sources with only two images available, the distance between the two reconstructed positions is reported.}\\
\hline
200 & 1 & $1.187 \pm 0.512$ \footnote{Source 200 has only one image. Its mean distance and uncertainty are computed based on all 12 delensed positions of Source 1.} & --- \\
2 & 7 & $0.380 \pm 0.145$ & $0.176$\\
3 & 6 & $0.585 \pm 0.185$ & $0.260$\\
4 & 7 & $0.941 \pm 0.117$ & $0.186$\\
5 & 2 & ($0.044, 0.039$) & $0.010$\\
7 & 5 & $1.005 \pm 0.302$ & $0.435$\\
8 & 4 & $0.528 \pm 0.127$ & $0.308$\\
12 & 2 & ($0.778, 0.777$) & $0.695$\\
14 & 2 & ($1.593, 1.737$) & $0.238$\\
16 & 2 & ($1.105, 0.780$) & $0.976$\\
18 & 2 & ($3.389, 2.646$) & $3.221$\\
19 & 2 & ($3.293, 2.769$) & $3.077$\\
51 & 2 & ($4.513, 4.249$) & $4.710$\\
101 & 5 & $1.567 \pm 0.420$ & $0.449$\\
\hline
\hline
\end{tabular}
\end{table}

\begin{figure*}
\centering
 \makebox[\textwidth]{\includegraphics[width=0.84\paperwidth]{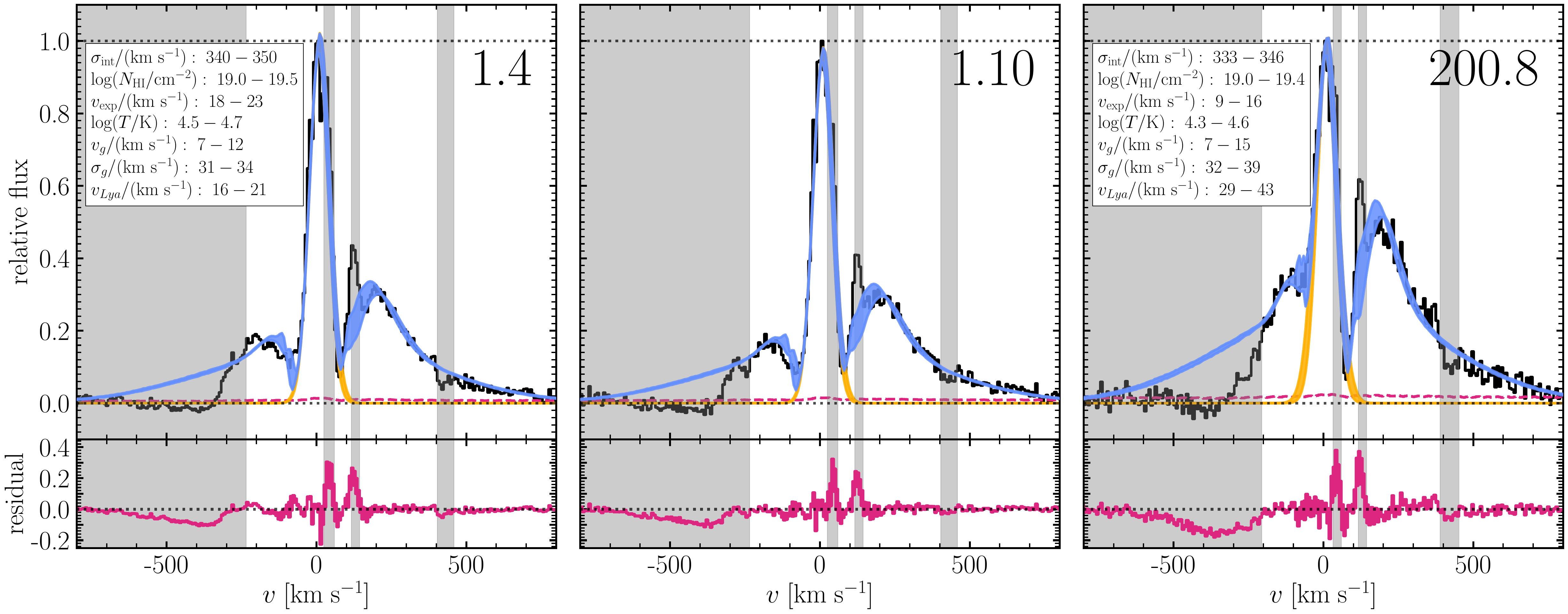}}
    \caption{Observed \lya\ profiles and the best-fit models of images 1.4, 1.10, and 200.8 in black. Following Figure \ref{fig:spectraoverview}, zero velocity corresponds to $z=2.37094$, the fiducial redshift of the LyC leaker \citep[e.g.,][]{Rivera-Thorsen2017}.  Recall that image 1.4 and 1.10 are two of the multiple images of the LyC leaker, while image 200.8 is a single image of a separate star-forming region at a projected distance of $\approx\!1.2$ kpc from the leaker.  A plausible range of best-fit parameters is determined by applying the mask iterations in Table \ref{tab:mask_ranges} when performing the maximum likelihood analysis (see Equation \ref{eqn:loglikelihood}). The LyC leaker profiles (1.4 and 1.10) are fitted together. The range of best-fit profiles from applying the mask iterations are plotted in sky-blue bands and central Gaussian components in orange bands. The \lya\ resonant scattering component is not separately plotted to prevent overcrowding. Masks are depicted in gray. Residuals between the data and the best-fit mode are presented in the bottom panel of each profile in magenta. These residuals suggest the presence of strong \lya\ absorbers at $\varv\!\approx\!-400$ \kms\ in the foreground IGM along the sightline between the observer and the Sunburst Arc.}
    \label{fig:fittingresults}
\end{figure*}

 \vspace{5pt}

\section{Spatial variation of \lya\ line in Sunburst}
\label{sec:results}

Following the descriptions presented in \S\ \ref{sec:lyaspectralanalysis}, we have performed a suite of radiative transfer simulations, utilizing a novel machine learning technique, to determine the underlying gas kinematics based on the observed \lya\ line profiles.  The results are summarized in Figure \ref{fig:fittingresults}.  As illustrated in \S\ \ref{sec:constraintsanduncertainties}, uncertainties in the best-fit model are driven by systematic uncertainties.  Therefore, a plausible range of best-fit parameters is presented here for the model profiles.

Recall that both Source 1, a LyC leaker, and Source 200, a LyC non-leaker, exhibit a similarly triple-peaked \lya\ profile, consisting of a Gaussian component sandwiched between an asymmetric double-peak component (see \S\ \ref{sec:quantifyinglyaspectraldifferences} and Figure \ref{fig:spectraoverview}). Our maximum likelihood analysis shows that the asymmetric double-peak component is best characterized by
an intrinsic source with a broad line width of $\sigma_{\rm int}\approx 330$--350 \kms.  This central powering source is embedded in a medium of neutral hydrogen column density $\log\,(N_{\rm HI}/\cmjj)\approx 19.0$--19.5 and effective temperature $\log\,(T/{\rm K})\approx 4.3$--4.7 for both the LyC leaker and non-leaker.  A notable difference is the velocity centroid of the intrinsic \lya\ powering source and the expansion velocity between the LyC leaker and non-leaker.  The LyC leaking region exhibits a relative velocity shift of $\varv_{\rm Ly\alpha}=+16$ to $+21$ from the fiducial redshift \citep[e.g.,][]{Rivera-Thorsen2017}\footnote{Note that uncertainties in the zero points of wavelength calibrations for different instruments affect the {\it absolute} redshift and velocity estimates of different sources, the effect of which is difficult to assess.  Therefore, we focus the discussion here on the {\it relative} velocities between components and between different regions based on the spectra acquired using the same instruments.}, and an expansion velocity of $\varv_{\rm exp}\approx 18$--23 \kms, while the LyC non-leaking "Godzilla" source exhibits a relative velocity shift of $\varv_{\rm Ly\alpha}=+29$ to $+43$ \kms\ and a more modest expansion velocity of $\varv_{\rm exp}\approx 9$--16 \kms.

In contrast, the Gaussian component is best characterized by a relatively narrow line width of $\sigma_g\approx 30$--40 \kms\ and velocity centroid of $\varv_g\approx +7$--15 \kms\ for both regions.  No significant velocity offset is found for the Gaussian component between the two sightlines.  This indicates that the scattered component is kinematically separated from the non-scattered component.

The residual spectra showing the difference between the observed and best-fit model profiles are also presented in Figure \ref{fig:fittingresults}. 
 Intriguingly, all three images share nearly identical residuals with negative (absorption-line like) fluctuations appearing at $\varv\!\lesssim\!-300$ \kms\ and $\varv\!\approx\!+400$ \kms\ and positive (emission-line like) fluctuations at $\varv\!\approx\!+40$ and $+130$ \kms, suggesting contributions from foreground objects.

A detailed gravitational lensing analysis has yielded a distance constraint of $1.2\!\pm\!0.5$ kpc between the two star-forming regions in the host ISM (see Table \ref{tab:separations}).  Allowing the possibility that Source 200 is part of Source 4, we still find a similar distance of $0.9\!\pm\!0.1$ kpc, consistent with $1.1_{-0.2}^{+0.7}$ kpc from \cite{Sharon2022}.

The difference in the observed \lya\ profiles between individual star-forming regions separated by merely $\sim\!1$ kpc in Sunburst highlights the uncertainties in interpreting integrated spectra from unresolved observations \citep[see also][]{Chen_2021}.  At the same time, the resemblance in the triple-peaked profile between LyC leaking and non-leaking regions also presents further challenges to the notion of utilizing the observed \lya\ line as a proxy for measuring LyC escape \citep[see also][]{Choustikov_2024a, Pahl:2024}.  Combining the \lya\ radiative transfer model analysis described in \S\ \ref{sec:lyaradiativetransfermodels} and the best-fit gravitational model described in \S\ \ref{sec:gravitationallensmodel}, here we describe the detailed spatial and spectral properties in this highly magnified starburst galaxy at {\it Cosmic Noon}.


\subsection{Kinematical Profile of the \lya\ Line}
\label{sec:kinematicallyresolvedlya}


Our \lya\ profile analysis shows that high-quality optical echelle spectra combined with the high magnification power of cluster lensing allow us to resolve ISM velocity structures on scales of $\approx\!10$ \kms\ between individual star-forming regions separated by $<\!1$ kpc.  Specifically, the individual star-forming regions, both the LyC leaker and non-leaker, are expanding at a similarly modest speed of $\varv_{\rm exp}\!\approx\!20$ \kms, and the non-leaking region (the "Godzilla") is redshifted from the leaker by $\Delta\,\varv_{\rm Ly\alpha}\!\approx\!+15$ \kms.  At the same time, both regions are surrounded by a more spatially extended ($\gtrsim\,1$ kpc) and highly ionized medium with $N_{\rm HI}\!\lesssim\!10^{13}\,\cmjj$ responsible for the observed common Gaussian component along both sightlines.

This relatively small line-of-sight velocity offset inferred between the two \lya\ scattering regions is consistent with the velocity offset observed using rest-frame optical nebular lines, such as [\ion{O}{3}]\,$\lambda\lambda\,4960, 5008$ or H$\alpha\,\lambda\,6564$, in recent {\it James Webb Space Telescope} ({\it JWST}) NIRSpec IFU observations \citep[e.g.,][]{Rivera:2024,Choe:2024}.  The agreement provides strong support for our \lya\ profile analysis.

At the same time, the inferred intrinsic line width of the scattered \lya\ component is large with $\sigma_{\rm int}\approx 340$ \kms, while the non-scattering Gaussian component shows a more modest width of $\sigma_g\approx 35$ \kms\ (or equivalently ${\rm FWHM}\approx 82$ \kms).  For comparison, optical nebular lines observed using {\it JWST} NIRSpec exhibit multi-component structures in both the LyC leaker \citep[see][]{Rivera:2024} and the non-leaker "Godzilla" \citep[see][]{Choe:2024}.  The observed narrower line width in the non-scattering \lya\ component agrees well with the line widths determined from optical nebular lines for both regions, while the broad intrinsic width of the scattering \lya\ photons is more in line with some of the broadest components seen in non-resonant lines.  However, whether the broad \lya\ wings and the broad component of non-resonant lines share a common origin remains unclear.  

The large line width required to explain the broad wings of scattered \lya\ photons under an expanding shell model is often seen in low-redshift green pea galaxies, which also display more modest line widths observed in non-resonant lines (e.g., \citealt{Orlitov:2018}; see also \citealt{Chen_2021} for star-forming galaxies at $z\approx 3$).  Such discrepancy between broad \lya\ wings and narrow non-resonant lines has raised questions about the validity of the adopted expanding shell models.  Recently, \cite{Li:2022} have shown that the inferred broad intrinsic line width can be attributed to clumpy gaseous structures that are not accounted for in a uniform expanding shell model, but the inferred total $N_{\rm HI}$, $T$, and $\varv_{\rm exp}$ remain robust in reflecting a global mean, irrespective of the assumed clumpiness in the models. This is qualitatively consistent with our findings of the two regions in Sunburst.  Despite the differences in the velocity centroids and the relative contribution of non-scattering \lya\ photons (see \S\ \ref{sec:nonleakercentrallyaflux} below), no significant variations in these quantities are found between the two star-forming regions separated by $\approx\!1$ kpc. 

\subsection{Significance of Central \lya\ Flux}
\label{sec:nonleakercentrallyaflux}

By targeting multiple sources within the Sunburst Arc—specifically, one exhibiting LyC leakage and another without—our study reveals a surprising finding: the presence of a central Gaussian peak in the non-leaking source. This triple-peak configuration, comprising a central Gaussian superimposed on an asymmetric two-peak \lya\ line, has been suggested as an indicator of ionized channels through which LyC photons can directly escape. At the same time, the cross-section of hydrogen atoms to \lya\ photons is $10^4$ times larger than LyC photons. Consequently, we anticipate detections of LyC photons from regions where direct escape of \lya\ photons is observed. However, we observe the opposite, namely the presence of directly escaped \lya\ with no LyC escape. 

Overall, the \lya\ profiles of the LyC leaker and non-leaker exhibit a triple-peaked shape. A closer quantitative comparison of the observed profiles reveals subtle morphological differences between the \lya\ profile of Source 200, the non-leaker, and that of Source 1, the leaker. The same quantitative comparison shows that the \lya\ profiles of images 1.4 and 1.10, two of the multiple images of Source 1, are statistically indistinguishable. We further quantify the difference based on the observed signal strengths 
by comparing the fractional \lya\ flux contained in the central Gaussian component for each image. We find that the central Gaussian component constitutes roughly $1/3$, $1/3$, and $1/4$ of the observed total \lya\ flux for images 1.4, 1.10, and 200.8, respectively. The observed central strength of the Gaussian component in the \lya\ line flux shows only a minimal difference between regions with LyC leakage and those without, with the central component contributing approximately $\approx 1/3$ more to the total flux in LyC leaker compared to the non-leaker. This minimal difference suggests that we should have expected to observe LyC leakage from the non-leaking regions as well.

\subsection{Contributions from Foreground Objects}
\label{sec:foregroundabsorbers}

The high resolving power in the MIKE spectra enables the identification of \ion{H}{1} absorption features that were previously unknown. We calculate the residual of our best-fit \lya\ models (see Figure \ref{fig:fittingresults}) and identify multiple absorbers along the line-of-sight that are coherent across all three sightlines. Among these are strong \lya\ absorbers located between $\varv\approx -500$ and $\varv\approx -350\,\mathrm{km} \, \mathrm{s}^{-1}$ relative to the Sunburst galaxy, consistent with the single absorption component noted by \cite{Rivera-Thorsen2017} in observations with the lower-resolution MagE spectrograph ($R \approx 4700$). These may be attributed to a cluster of \lya\ forest absorbers 
at $z\approx 2.365$ and 2.367.  At these redshifts, the three sightlines are separated by between 0.9 and 1.7 kpc in the source plane.  The resemblance between the residual absorption spectra of the three sightlines therefore indicates that the IGM is highly coherent on scales of 1-2 kpc \citep[see also][]{Rauch:2001}.

In addition, large residuals are seen at $\varv\approx 0$--150 \kms. Interpreting these features is not straightforward, especially in the presence of the IGM \lya\ forest absorption.  Specifically, the emission-line peaks at $\varv\,\approx\,+40$ and $+130$ \kms\ along all three sightlines may result from foreground \lya\ absorption, masquerading as emission features in the residual spectra. Alternatively, the symmetric profiles may originate in isolated star-forming regions in Sunburst. Incidentally, recent {\it JWST} NIRSpec observations of both of these sources have also revealed multi-components structures in [\ion{O}{3}] and H$\alpha$ \citep[e.g.,][]{Rivera:2024,Choe:2024}, but establishing one-to-one correspondence of individual components between different transitions remains challenging.



\begin{table*}
\renewcommand{\arraystretch}{1.5}
\centering
\caption{Range of best-fit parameters for the \lya\ profiles in the three images $1.4$, $1.10$, and $200.8$ in the Sunburst Arc, corresponding to the range of masks applied (see \S\ \ref{sec:constraintsanduncertainties}).} 
\begin{tabular}{c|c|c|c|c|c|c|c}
\hline
\hline
 & $\sigma_{\mathrm{int}} / (\mathrm{km} \, \mathrm{s}^{-1})$ & $\mathrm{log}(N_{\mathrm{HI}} / \mathrm{cm}^{-2})$ & $\varv_{\mathrm{exp}} / (\mathrm{km} \, \mathrm{s}^{-1})$ & $\mathrm{log}(T \, / \, \mathrm{K})$ & $\varv_{g} / (\mathrm{km} \, \mathrm{s}^{-1})$ & $\sigma_{g} / (\mathrm{km} \, \mathrm{s}^{-1})$ & $\varv_{\mathrm{Ly\alpha}} / (\mathrm{km} \, \mathrm{s}^{-1})$ \\
\hline
1.4 and 1.10 & 340 - 350 & 19.0 - 19.5 & 18 - 23 & 4.5 - 4.7 & 7 - 12 & 31 - 34 & 16 - 21 \\
200.8 & 333 - 346 & 19.0 - 19.4 & 9 - 16 & 4.3 - 4.6 & 7 - 15 & 32 - 39 & 29 - 43 \\
\hline
\hline
\end{tabular}
\label{tab:results}
\end{table*}

\section{Discussion}
\label{sec:discussion}

This work utilizes a combination of strong gravitational lensing and \lya\ radiative transfer to resolve the ISM properties of distant star-forming galaxies on sub-kpc scales \citep[see also e.g.,][]{Chen_2021}. We have analyzed the \lya\ emission profiles of two distinct star-forming regions in the highly magnified Sunburst Arc galaxy. This galaxy is unique because of its bright LyC leakage detected at one location but is otherwise a typical starburst galaxy at cosmic noon. The presence of a foreground galaxy cluster provides a strong gravitational lensing power to produce extended arcs around the cluster's Einstein radius and enable spatially resolved maps of individual star-forming clumps. The combined imaging and spectroscopic data allow us to examine the variations in gas properties between clumps on scales of a few 100 parsecs, 
which would otherwise be impossible for galaxies at cosmic noon, $z \sim 2$. 
In this section, we discuss the implications of our work on the connection between \lya\ and LyC photon escape in reionization analog galaxies.

\subsection{Connecting LyC Leakage to \lya\ Profiles}
\label{sec:discussion_LyCescape}

The Sunburst Arc presents a unique opportunity to investigate the relationship between LyC and \lya\ emission. Alongside the typical scattered \lya\ observed in many \lya\ emitters \citep[e.g.,][]{Izotov_2018c, Flury_2022a, Flury_2022b}, the presence of a central peak of non-scattered \lya\ is especially notable in this system. While both LyC and \lya\ escape routes likely share similarities due to their dependence on low neutral hydrogen optical depth, their vastly different opacities complicate understanding their mutual influence. It is much harder for \lya\ to escape through neutral hydrogen than LyC, so in any case where there is non-scattered \lyacom\ one would naturally expect LyC leakage. Nonetheless, this simple picture is not consistent with our \lya\ spectra. There is a clear central Gaussian of non-scattered \lyacom\ which is only diminished by $\sim 1/3$ in the non-leaker compared to the LyC leaker (see \S\ \ref{sec:nonleakercentrallyaflux}). 

In our analysis, we consider the possibility that the scattered \lya\ signals are powered by \textit{in situ} star formation within a thin shell of neutral gas, effectively a "Strömgren sphere" \citep{Stromgren_1939}. From this model, we determine the \ion{H}{1} column density of the gas shell surrounding both star-forming regions (leaker and non-leaker) to be between $19.0 < \mathrm{log}(N_{\mathrm{HI}}/\mathrm{cm}^{-2}) < 19.5$ which confidently implies that any such shell should scatter \lya\ photons. 
While this column density is sufficient to trap LyC photons in both regions, it is somewhat peculiar that the leaker and non-leaker, separated by $\sim 1 \, \mathrm{kpc}$, fall within the same range of \ion{H}{1} column densities, given that only one region exhibits LyC leakage. This suggests that additional factors, such as the presence of clear sightlines free of neutral hydrogen gas in order to let through the \lya\ and LyC photons. Such clear sightlines has been invoked to support the idea of anisotropic LyC escape, which can increase the escape fraction to $f_{\mathrm{esc, LyC}} \gtrsim 20\%$ required to complete reionization \citep[e.g.,][]{Robertson_2015, Faucher-Giguere_2020}, and has motivated perforated shell models like the one used in this work.


In addition, by employing high-resolution spectroscopy, we have determined the relative kinematics between the scattered and non-scattered \lya\ components to narrow down their possible origins. As mentioned in \S\ \ref{sec:results}, \lya\ has four orders of magnitude larger opacity than LyC, so one would expect to see significantly less central \lya\ flux in non-leakers due to the increased scattering and absorption by neutral hydrogen. Indeed, previous studies have utilized the peak separation, a proxy for the underlying $N_{\rm HI}$, to infer $f_{\mathrm{esc, LyC}}$, finding that $f_{\mathrm{esc, LyC}} = 3.23 \times 10^4 (\varv_{\mathrm{sep}} / \kms)^{-2} - 1.05 \times 10^2 (\varv_{\mathrm{sep}} / \kms)^{-1} + 0.095$ \citep[e.g.,][]{Izotov_2018c}.  Applying this relation, we would infer $f_{\mathrm{esc, LyC}}\sim 5\%$ for the LyC leaker and $\sim 10\%$ for the non-leaker.  This is inconsistent with the observed LyC escape of  $\gtrsim 50\%$ from the Sunburst Arc in the F275W band \citep[see][]{Rivera-Thorsen2019}.


At the same time, these inconsistencies that occur with more complex profiles (e.g., triple-peaked) motivated \cite{Naidu_2022} to introduce the \textit{central escape fraction}, $f_{\mathrm{cen}}$, defined by 
\begin{equation}
    \label{eqn:f_cen}
    f_{\mathrm{cen}} = \frac{\mathrm{Ly}\alpha \; \mathrm{flux \; within} \; \pm 100 \; \mathrm{km \; s^{-1}}}{\mathrm{Ly}\alpha \; \mathrm{flux \; within} \; \pm 1000 \; \mathrm{km \; s^{-1}}}.
\end{equation}

Both the LyC leaker and the non-leaker in this work exhibit high $f_{\mathrm{cen}}$ of $42\%$ and $37\%$, respectively. Since the differences in central \lya\ is small between the LyC leaker and non-leaker within the Sunburst Arc, our work argues that central \lya\ flux alone is not a good diagnostic for LyC escape \citep[see also][]{Choustikov_2024a}.

From our computed gravitational lens model, the \lya\ kinematics vary locally on small scales of $\lesssim 1 \, \mathrm{kpc}$. Yet, the central Gaussian centroid velocity remains unchanged from the LyC leaking clump (Source 1) to the non-leaking clump (Source 200). One interesting consequence of comparing these relative kinematics is that we can examine whether escaping LyC photons originate in the same location as the directly escaping \lya\ photons. While our simple shell model successfully reproduces the scattered \lya, the centroid of the scattered \lya\ component in the non-leaking region is redshifted relative to the LyC leaker by $\Delta\,\varv_{\rm Ly\alpha}\!\approx\!+15$ \kms, 
implying a velocity shear between the two regions. Yet, the velocity centroid of the Gaussian is consistent across the sources, 
suggesting that a significant amount of the non-scattering \lya\ photons are not powered by the same source. If we also consider the large spatial separation between the two regions, this implies the presence of a global ($\gtrsim 1 \, \mathrm{kpc}$) source of \lya\ undergoing no scattering. In nearby galaxies, \ion{H}{2} regions are ubiquitous, providing plenty of \lya\ photons from recombination.

As illustrated in Figure \ref{fig:cartoon}, the Sunburst Arc can have a volume-filling warm ionized halo providing the bulk of the non-scattered \lyaper\ As long as the density of this halo is low, the ionizing source can be a few stars within the star-forming regions, such as the very massive stars recently proposed in the LyC leaker by \cite{Mestric_2023}, or be the ultraviolet background (UVB). This also provides a means for \lya\ emission without scattering that is independent of LyC escape being produced \textit{in situ} within the star-forming clumps. A commonly considered simple picture of LyC leakage involves two primary modes of escape: (i) \textit{Ionization-bound LyC leakage} where LyC photons escape through holes or chimneys of low neutral column density in the ISM; and (ii) \textit{Density-bound LyC leakage} where LyC production from massive stars is strong enough to ionize its way through all the gas in the system \citep[see e.g.,][]{Zackrisson_2013}. However, since the LyC and central \lya\ do not share the same origin, it is necessary to consider a different scenario that allows 
the non-scattered \lya\ photons to escape. 
A natural explanation may be a volume-filling ionized gas in an \ion{H}{2} region producing a clear sightline of \lya\ through recombination. The maximum neutral hydrogen column density of such an ionized halo cannot exceed $N_{\rm HI}=10^{13} \, \mathrm{cm}^{-2}$ where the \lya\ optical depth, $\tau_{\mathrm{Ly\alpha}}$, becomes greater than $\sim 1$ as this would scatter the \lya\ photons away from the line-of-sight. The size of this halo, $d$, and its neutral hydrogen density, $n_{\mathrm{HI}}$, are therefore governed by 
\begin{equation}
\label{eqn:tau_lya}
    \tau_{\mathrm{Ly\alpha}} \sim \left( \frac{n_{\mathrm{HI}}}{10^{-10} \, \mathrm{cm}^{-3}} \right) \, \left( \frac{d}{100 \, \mathrm{kpc}} \right) < 1.
\end{equation}

\begin{figure*}
\centering
{\includegraphics[width=0.825\paperwidth]{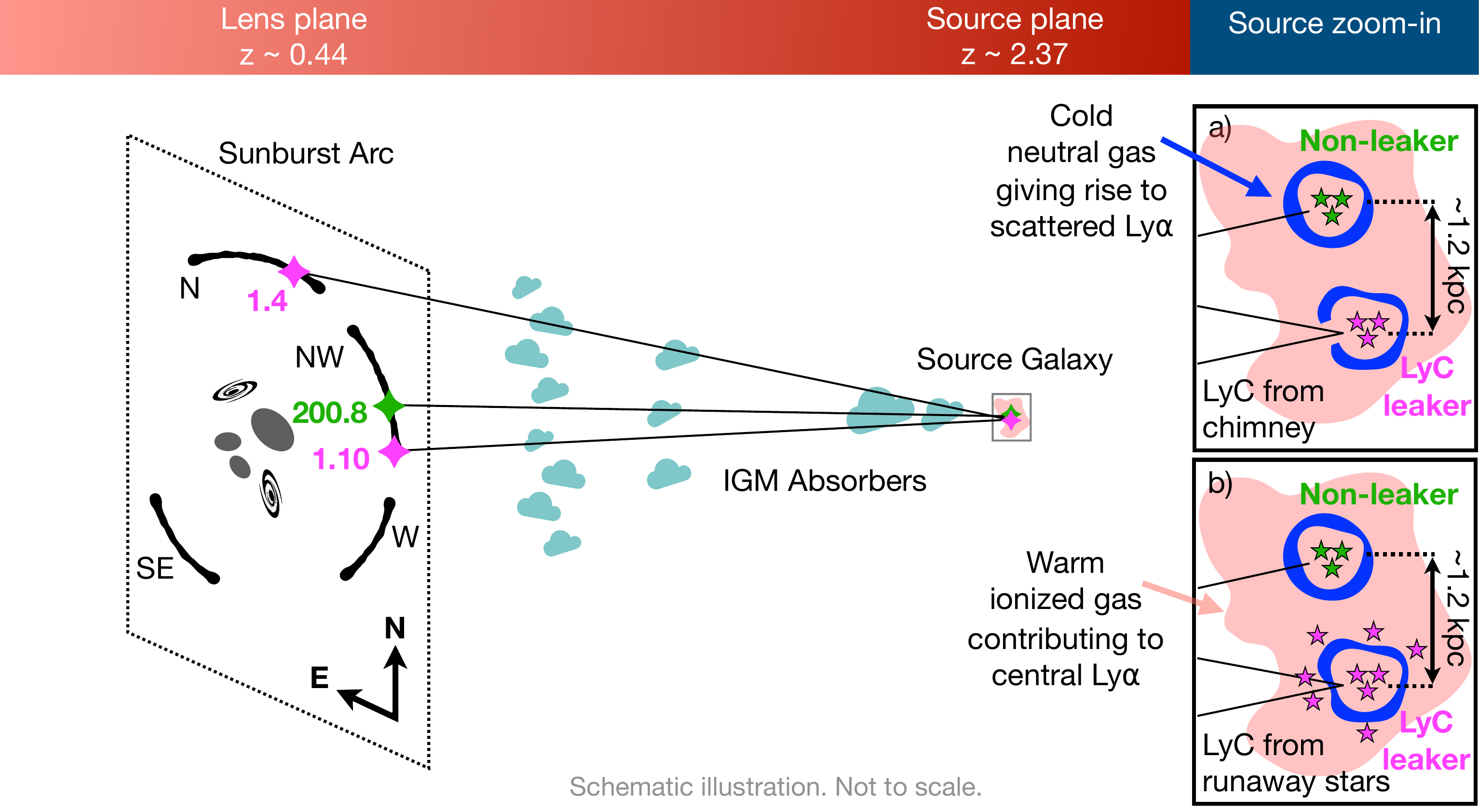}}
    \caption{Illustration depicting the escape of LyC and \lya\ photons from resolved clumps in the Sunburst galaxy, which is gravitationally magnified by a foreground galaxy cluster at $z\approx 0.44$. Our spectral analysis has uncovered kinematic variations across a spatial separation of approximately $1 \, \mathrm{kpc}$ in the source plane. The kinematic properties of the observed \lya\ profiles between different regions argue for distinct origins between directly escaped \lya\ and LyC photons. 
    While the directly escaping \lya\ photons arise in a spatially extended, warm ionized medium (highlighted in pale red), an expanding ionization-bounded shell (in blue) surrounding young stars produces the asymmetric double-peak \lya\ line. The IGM absorbers closest to the source galaxy in the illustration represent the absorbers between $\varv\approx -500$ and $\varv\approx -350\,\mathrm{km} \, \mathrm{s}^{-1}$ relative to the Sunburst galaxy.
    Two scenarios are proposed to explain the production and escape of LyC photons. Scenario a) involves a clear ionized channel directed towards the observer with a low density of neutral hydrogen, while Scenario b) includes an ensemble of runaway stars emerging from the surrounding gas providing a low-column density pathway for LyC escape.}
    \label{fig:cartoon}
\end{figure*}

Figure \ref{fig:cartoon} illustrates the geometric configuration between different escaping channels. 
First, a warm ionized medium produces the directly escaping \lya\ photons, while an expanding ionization-bounded shell surrounding young stars produces the asymmetric double-peak \lya\ line. From this, there are two possible scenarios for LyC escape: a) through ionized holes or "chimneys" (top right panel of Figure \ref{fig:cartoon}), or b) from runaway stars with a transparent line-of-sight (bottom right panel of Figure \ref{fig:cartoon}). While the former scenario is commonly seen in models of anisotropic LyC escape \citep[e.g][]{Choustikov_2024a}, runaway stars at high-velocity tails \citep[e.g.,][]{Blaauw:1961, Gies:1986} provide a likely alternative for the source of ionizing photons beyond their birth site \citep[e.g.,][]{Conroy:2012}. This hypothesis entails that many stars undergo a significant increase in velocity, due to collisions in star clusters and events like mergers or tidal interactions, which allows them to peek through the neutral ISM. Such runaway stars are observed in nearby disk galaxies \citep[e.g.][]{Hills_1988, Brown_2005}. These stars can fuel the diffuse warm ionized halo sourcing the \lya\ along with the ultraviolet background. At present, we are unable to rule out either of these scenarios.

While our work suggests that $f_{\mathrm{cen}}$ may be an incomplete indicator of LyC escape, it can still trace low-opacity channels. As noted by \cite{Choustikov_2024b}, the non-leaker could have a high escape fraction but low UV photon production, implying that the non-leaking stellar population must be significantly old enough to have passed its UV-strong phase experienced during the first $\sim 5$ Myr. \cite{Mainali_2022} report ages of $3.3 \pm 0.5$ Myr 
for the leaking region, and $11.8 \pm 0.9$ Myr 
for a stack of non-leaking regions in the Sunburst Arc, including Source 200. Since UV brightness is at its strongest in the first few Myr, it is possible that the escape fraction is high in the non-leaker but the stellar population has surpassed its most UV emissive era. However, note that the age estimate for the non-leaker is stacked for several non-leakers, not individually for Source 200, further complicating the interpretations.

Our results support the picture that the Sunburst galaxy is a dynamic system of multiple star-forming regions with dense clouds flowing outwards at a modest speed ($\varv_{\rm exp}\approx 20$ \kms) from the star-forming regions. Both the LyC leaker and non-leaker provide unambiguous examples of gas outflows which are believed to be ubiquitous at high redshifts and necessary to provide metal enrichment of the IGM \citep[e.g.,][]{Aguirre_2001, Oppenheimer_2006, Booth_2012} and to regulate galaxy growth \citep[e.g.,][]{Hopkins_2014}. The properties derived from the \lya\ profiles in this study are characteristic of the interstellar medium (ISM) of star-forming galaxies, featuring outflow velocities of approximately $\sim 20 \, \mathrm{km} \, \mathrm{s}^{-1}$ and temperatures of $\sim 10^{4.5} \mathrm{K}$. These mild stellar winds derived from the \lya\ peak ratios 
are nonetheless insufficient to shift the \lya\ rest frame to facilitate direct escape of \lya\ from the optically thick shell.
We therefore conclude that the escaping LyC photons, which are produced in stellar populations together with the scattered \lya\ photons, are not generated at the same site as the directly escaping \lya\ photons.

\subsection{The Nature of the Exceptionally Bright Non-Leaker}
\label{sec:godzilla}

The non-leaking source analyzed in this work (Source 200, or "Godzilla") exhibits exceptional brightness which sets it apart from the other non-leaking clumps in the Sunburst Arc. Understanding the non-leaker's physical origin is useful in interpreting the stellar populations sourcing its \lya\ signature. Previous work has suggested that the non-leaker could be a transient event in which the source is sufficiently small and crossing the source-plane caustic \citep{Vanzella_2020, Diego_2022}. However, \cite{Sharon2022} showed that this was inconsistent with the expected time delay between the N and NW arcs. Recently, \cite{Pascale_2024} suggested Godzilla is a super star cluster surrounded by ejecta from massive stars, evidenced by metal enrichment typical of core-collapse supernovae.

In this work, we provide additional constraints on the nature of the non-leaker. We show that the non-leaker, (while having negligible contamination by nearby 300.80 and 300.81) shows significant central Gaussian flux in the rest frame of \lyacom\ confidently placing it at the same redshift as the Sunburst Arc source galaxy ($z = 2.37$) and ruling out the possibility of the non-leaker being contamination from a foreground cluster galaxy. While its color is not consistent with being a multiple image of Source 4, our lens model predicts an additional Source 4 image $\sim 0.5''$ east of the observed location of non-leaker with a magnification $3-4$ times lower than the pair. By itself, our lens model allows for the non-leaker to be part of Source 4. However, there is a large uncertainty in the magnification factor between these predicted images given their proximity to the critical curve, so slight changes in the model can remove this additional image prediction for Source 4.

Previous discussions of the non-leaker have revolved around the lack of multiple images. Using our lens model, we can calculate a lower limit of the intrinsic brightness of the non-leaker. First, we perform a ray tracing of the non-leaker to the source plane and perform a relensing with \texttt{GLAFIC} to produce the locations of predicted images of the non-leaker (see green crosses in Figure \ref{fig:lens}). Our lens model predicts four other images of the non-leaker. If the magnification factors of the observed non-leaker image are significantly higher than the other predicted images, which is a reasonable expectation given the proximity of the non-leaker to the caustic, it is possible we only detect one of its images (image 200.8). By measuring the flux within the FWHM of image 200.8 in the F606W band, we derive an apparent AB magnitude of $m_{\mathrm{F606W}} \sim 22.3$. The magnification field from our lens model yields a magnification factor $\mu \sim 4000$ at the centroid location of the non-leaker. By taking the ratios of the magnification factor of 200.8 and the lowest magnification of the four predicted images, we obtain a lower limit for the apparent magnitude of $m_{\mathrm{F606W}} \sim 28.2$ for these additional images. This is comparable to the background brightness $m_{\mathrm{F606W}} \sim 28.0$ estimated by moving the aperture slightly off the target to a dark patch of the sky. The delensed (intrinsic) magnitude of the non-leaker is thus $M_{\mathrm{F606W}} \geq -15$. This supports the idea that the non-leaker can be a compact star-forming region located at the caustic, making its other images undetectable. However, note that there is a large uncertainty in the magnification factor of image 200.8, and a magnification of $\sim 4000$ implies that the source cannot be larger than 0.2 pc in diameter assuming tangential and radial magnifications of $\sim 2000$ and $\sim 2$; otherwise it would have been resolved in \textit{HST} and \textit{JWST} images. This poses a challenge, as no known compact group of stars could provide the observed luminosity within such a small volume.

\section{Summary and Conclusions}
\label{sec:summary}

The \lya\ emission line, through its resonant scattering with neutral hydrogen, is a powerful probe of high-redshift galaxies capable of revealing potential channels of LyC escape. However, analyzing \lya\ comes with caveats, namely the need for careful modeling of its radiative transfer and the large difference in opacity between \lya\ and LyC ($\sim10^4$). Here, we show an example demonstrating that the central \lya\ flux method is not a reliable standalone indicator of LyC leakage, and a direct implication of our findings is that measurements based on integrated light should be taken with caution.

We present the first high-resolution ($R \approx 29{,}000$) spectral analysis of the \lya\ profiles of two distinct star-forming regions within the Sunburst Arc: one leaking ionizing photons (Source 1, dubbed the "LyC leaker"), and one without observed ionizing photon escape (Source 200, nicknamed the "non-leaker"). We obtain ground-based optical spectra with the Magellan MIKE spectrograph. One of the primary goals of our analysis is to determine the spatial separation between the LyC leaker and non-leaker in the source plane to aid our understanding of how LyC photons escape the Sunburst Arc. We create a lens model of the foreground mass distribution to produce a quantitative reconstruction of the star-forming clumps in the Sunburst Arc. 

Our analysis focuses on the spectral diagnostics of the \lya\ profiles of three knots, two being multiple images of the LyC leaker (images 1.4 and 1.10) and one being an exceptionally bright non-leaker with a puzzling origin due to its single image along the arc and its high magnification. We unsurprisingly find that the \lya\ profiles of the two LyC leaking images are statistically indistinguishable from each other. Although the leaker's profile is distinctly different from that of the non-leaker, both exhibit a notable triple-peaked configuration, which is commonly associated with LyC leakage. This is surprising because we would not expect the non-leaker to show such a configuration if it indeed lacks LyC leakage. The fact that the non-leaker exhibits a central peak suggests that the non-scattered \lya\ is not produced \textit{in situ} along with the LyC, and we further confirm this by comparing the relative kinematics between the non-scattered and scattered \lya\ components in each of the three spectra. The velocity centroids of the central Gaussian \lya\ component remain unchanged across all sources, while those of the scattered \lya\ component vary from leaker to non-leaker, consistent with our assumption that these \lya\ components originate from distinct physical locations. Additionally, the relative strengths of the central Gaussian and resonantly scattered components is diminished in the non-leaker by only $\sim 1/3$. Our analysis suggests the origin of the directly escaping \lya\ is a volume-filling, warm ionized medium spanning $\sim\!1$ kpc and that the central \lya\ does not necessarily trace directly escaping LyC photons.

The high spectral resolution of our MIKE spectra is essential for resolving the kinematic differences between the LyC leaker and non-leaker \lya\ profiles and the multi-component foreground absorption, which is coherent across all spectra. This resolution allows precise measurement of the velocity centroids of both the central Gaussian and scattered \lya\ components, revealing their distinct physical origins. In contrast, previous studies using lower resolution MagE spectra ($R \approx 4700$) \citep{Rivera-Thorsen2017, Mainali_2022, Owens_2024}, lacked the spectral detail needed to fully resolve the kinematics of the \lya\ profiles. Our MIKE spectra also reveals multiple \lya\ absorptions along the line-of-sight. An especially strong \lya\ multi-component absorber at roughly $-400 \, \mathrm{km} \, \mathrm{s}^{-1}$ relative to the \lya\ rest frame implies the potential presence of a large reservoir of infalling neutral gas or foreground IGM absorption. The presence of this absorption feature in all spectra supports that the IGM is coherent across scales of $1 - 2$ kpc.

Our fiducial gravitational lens model predicts a distance between the LyC leaking clump (Source 1) and the non-leaking clump (Source 200) of $\sim 1.2 \, \mathrm{kpc}$, consistent with previous studies, which sets the lower limit for the size of the warm ionized halo sourcing non-scattered \lya. Since the exceptionally bright non-leaker is not multiply imaged, we investigate the possibility that a low intrinsic brightness and extreme magnification can cause its singly imaged nature. We relens the non-leaker and calculate a lower limit for the intrinsic brightness of the non-leaker and find that it has $M_{\mathrm{F606W}} \geq -15$ and an apparent brightness lower than the sky background.

In conclusion, our analysis of \lya\ profiles and gravitational lens modeling provides valuable insights into the properties of the observed sources and the intervening gas along the line-of-sight. The hydrogen \lya\ line is the strongest emission line in diffuse, photoionized gas and enables sensitive studies of outflows from star-forming regions. 
Yet, the uncertainties in the outflows are large given the intervening absorption by neutral hydrogen and the uncertainty in the geometry needed to model \lya\ radiative transfer. 
To elucidate the connection between \lya\ and LyC in reionization analogs, both spatial and spectral resolution is needed in order to accurately model the \lya\ profiles. Therefore, we recommend quantitative studies of lensed \lya\ arcs for observational studies of the ionizing photon budget in the early universe.

\section*{Acknowledgments}
\label{sec:acknowledgments}
The authors thank Jose Diego for helpful comments on an earlier version of the paper, Rui Marques-Chaves for pointing out relevant references on detections of Lyman continuum leakers, Alex Ji for pointing us to the neural network architecture \texttt{ThePayne}, Yuan-Sen Ting, the author of \texttt{ThePayne}, for stimulating discussions on the application of the code, and Ava Polzin for assistance in setting up and testing the GitHub repository of \texttt{CCLya-Payne}.  H.-W.C., E.S., and M.C.\ acknowledge partial support from HST-GO-17146.004A and The University of Chicago Women's Board grants. E.S. acknowledges support from the Brinson Foundation.  M.C. acknowledges support from the University of Chicago Data Science Institute (DSI). M.G. thanks the Max Planck Society for support through the Max Planck Research Group. S.L. acknowledges support by FONDECYT grant 1231187. This work was completed with resources provided by the University of Chicago Research Computing Center.

This paper includes data collected with the 6.5 m Magellan Telescopes located at Las Campanas Observatory (LCO), Chile. The Magellan MIKE observations were carried out as part of programs CN2019A-48 (Chile CNTAC) and 2019A-02 (Chicago TAC).
The authors extend special thanks to the LCO staff for their expert support. 
This work has made use of data from the European Space Agency (ESA) mission
{\it Gaia} (\url{https://www.cosmos.esa.int/gaia}), processed by the {\it Gaia} Data Processing and Analysis Consortium (DPAC, \url{https://www.cosmos.esa.int/web/gaia/dpac/consortium}). Funding for the DPAC has been provided by national institutions, in particular the institutions participating in the {\it Gaia} Multilateral Agreement.

\textit{Software:} \texttt{astropy} \citep{Astropy_2013, Astropy_2018}, \texttt{DS9} \citep{Smithsonian_2000}, \texttt{emcee} \citep{Foreman-Mackey_2013}, \texttt{GLAFIC} \citep{Oguri_2010}, \texttt{IRAF} \citep{Tody_1986, Tody_1993}, \texttt{matplotlib} \citep{Hunter_2007}, \texttt{numpy} \citep{Harris_2020}, \texttt{scipy} \citep{Virtanen_2020}, \texttt{Source Extractor} \citep{Bertin_1996}, \texttt{TLAC} \citep{Gronke_2014}.

\appendix{}

\section{Image Constraints for Lens Model}
\label{appx:image_constraints}

To measure the positions of the constraints that go into our lens model, we stack \textit{HST} F606W, F814W, and F160W images. A color RGB image compiled with these is provided in Figures \ref{fig:spectralocations} and \ref{fig:lens}. We use the knot identifications from \cite{Sharon2022} and independently measure the centroids of each knot in our stacked F814W image with the image processing tool \texttt{IRAF} \citep{Tody_1986, Tody_1993}.

Image constraints are determined by their pixel centroids in the F814W stacked image and translated into $\Delta \mathrm{RA}$ and $\Delta \mathrm{DEC}$ given by

\begin{equation}
\label{eqn:deltaRA}
    \Delta \mathrm{RA} = (\mathrm{RA} - \mathrm{RA}_{\mathrm{BCG}}) \cdot \mathrm{cos}(\mathrm{DEC}_{\mathrm{BCG}}) \cdot (-1)
\end{equation}
\begin{equation}
\label{eqn:deltaDEC}
    \Delta \mathrm{DEC} = \mathrm{DEC} - \mathrm{DEC}_{\mathrm{BCG}}
\end{equation}
where $(\mathrm{RA}_{\mathrm{BCG}}, \mathrm{DEC}_{\mathrm{BCG}}) = (15{:}50{:}07.115, \, -78{:}11{:}29.953)$ is the right ascension and declination coordinate of the central BCG determined by measuring the centroid of the most central galaxy in the central cluster (contamination of the central cluster galaxies by foreground stars makes measuring the optical centroid of the BCG challenging). The factor of $-1$ comes from the fact that \texttt{GLAFIC} treats West as the $+x$-direction. North is considered the $+y$-direction.

The Sunburst Arc contains a total of 12 individual images. Source 1 is the only source that is multiply imaged 12 times while the rest of the sources have varying image multiplicities between one and 12. The name of each knot contains information about the source and its location in the lens configuration, e.g. "1.8" indicates that the knot corresponds to Source 1 and is located in image 8 in the lens/image plane. Below, we provide a summary of the sources making up our list of image constraints:

\begin{itemize}
\item{\textit{Source 1, 2, 3, 4, 5, 7, 8, 12, 14, 16, 18, 19, 51, 52, 101, and 300.} These sources are multiply imaged star-forming clumps within the primary galaxy of the Sunburst Arc at $z \approx 2.37$. We treat Source 300 (the "Pair") as part of Source 4.}

\item{\textit{Source 41.} This source is located between the foreground cluster and the Sunburst galaxy at a confirmed spectroscopic redshift of $z=1.1860$ \citep{Sharon2022}. While it is located in a region that theoretically contributes to the observed lensing potential, we choose not to model it as a halo contributing to the total mass distribution of the model due to its predicted location being near the center of the lens and thus having insignificant effect on the lensing potential.}

\item{\textit{Source 93, 94, and 95.} These sources are part of the same system, a galaxy behind Sunburst with a spectroscopic redshift of $z=3.5100$ \citep{Sharon2022}.}
\end{itemize}

We choose not to include sources that lack spectroscopically determined redshifts and images that are too faint to determine their centroid in the stacked images. We also do not include sources for which we can confidently measure just one of its images, as they do not provide constraining power for the lens model. Source 200 is not included as a constraint due to its single image multiplicity. A list of the measured coordinates for all image constraints used in this work is provided in Table \ref{tab:image_constraints}.

\begin{table}
\renewcommand{\arraystretch}{1.2}
\centering
\caption{Image constraints. $(\Delta \mathrm{RA}, \, \Delta \mathrm{DEC})$ measured relative to $(\mathrm{RA}_{\mathrm{BCG}}, \, \mathrm{DEC}_{\mathrm{BCG}}) = (15{:}50{:}07.115, \, -78{:}11{:}29.953) $. }
\label{tab:image_constraints}
\begin{tabular}{l|c|c|c|p{2cm}}
\hline
\hline
Object ID & $\Delta$ RA [$"$] & $\Delta$ DEC [$"$] & $z_{\mathrm{spec}}$ & Notes \\
\hline
\hline
1.1 & $-$0.9155 & 32.7405 & 2.3709 & Sunburst main galaxy \\
1.2 & 2.95 & 31.8675 &  &  \\
1.3 & 3.45 & 31.65 &  &  \\
1.4 & 7.7 & 30.35 &  &  \\
1.5 & 8.35 & 30.066 &  &  \\
1.6 & 8.916 & 29.7135 &  &  \\
1.7 & 15.0 & 25.05 &  &  \\
1.8 & 20.65 & 19.274 &  &  \\
1.9 & 21.931 & 17.5095 &  &  \\
1.10 & 22.556 & 16.329 &  &  \\
1.11 & 26.276 & 3.058 &  &  \\
1.12 & $-$24.5 & $-$17.47 &  &  \\
\hline
2.2 & 1.595 & 32.3085 & 2.3709 & \dittoclosing \\
2.3 & 4.1715 & 31.35 &  &  \\
2.4 & 6.95 & 30.6415 &  &  \\
2.7 & 15.3 & 24.7 &  &  \\
2.8 & 18.984 & 21.183 &  &  \\
2.11 & 26.2065 & 2.9 &  &  \\
2.12 & $-$24.8 & $-$17.15 &  &  \\
\hline
3.3 & 4.485 & 31.25 & 2.3709 & \dittoclosing \\
3.4 & 6.55 & 30.7755 &  &  \\
3.7 & 15.661 & 24.3675 &  &  \\
3.8 & 18.1695 & 22.0 &  &  \\
3.11 & 26.1985 & 2.719 &  &  \\
3.12 & $-$24.95 & $-$16.9415 &  &  \\
\hline
4.4 & 6.55 & 30.6185 & 2.3709 & \dittoclosing \\
4.7 & 15.117 & 24.72 &  &  \\
300.80 & 21.8 & 17.4 &  &  \\
300.81 & 26.15 & 3.4 &  &  \\
4.9 & $-$24.6825 & $-$17.431 &  &  \\
4.11 & 20.0245 & 19.7895 &  &  \\
4.12 & 20.1 & 19.6825 &  &  \\
\hline
5.1 & $-$0.6795 & 32.7 & 2.3709 & \dittoclosing \\
5.2 & 2.7175 & 31.9575 &  &  \\
\hline
7.1 & 0.2 & 32.837 & 2.3709 & \dittoclosing \\
7.2 & 1.854 & 32.5295 &  &  \\
7.7 & 16.271 & 24.169 &  &  \\
7.8 & 17.387 & 23.1 &  &  \\
7.11 & 26.2 & 0.8165 &  &  \\
\hline
8.1 & 0.7 & 32.5715 & 2.3709 & \dittoclosing \\
8.2 & 1.788 & 32.3495 &  &  \\
8.8 & 18.538 & 21.7385 &  &  \\
8.11 & 26.24 & 2.388 &  &  \\
\hline
12.3 & 5.0235 & 31.22 & 2.3709 & \dittoclosing \\
12.4 & 5.95 & 31.02 &  &  \\
\hline
14.7 & 16.3235 & 23.574 & 2.3709 & \dittoclosing \\
14.8 & 17.05 & 22.8895 &  &  \\
\hline
16.11 & 26.15 & 1.7 & 2.3709 & \dittoclosing \\
16.12 & $-$25.2495 & $-$16.3755 &  &  \\
\hline
18.11 & 25.65 & $-$1.6145 & 2.3709 & \dittoclosing \\
18.12 & $-$26.687 & $-$13.9 &  &  \\
\hline
\hline
\end{tabular}
\end{table}
\begin{table}
\renewcommand{\arraystretch}{1.2}

\centering
\caption{Image constraints (continued).}
\begin{tabular}{l|c|c|c|p{2cm}}
\hline
\hline
19.11 & 25.6645 & $-$1.8715 & 2.3709 & Sunburst main galaxy \\
19.12 & $-$26.815 & $-$13.7 &  &  \\
\hline
51.11 & 25.25 & $-$3.677 & 2.3709 & Associated with Sunburst \\
51.12 & $-$27.65 & $-$11.7395 &  &  \\
\hline
101.1 & $-$2.3075 & 33.272 & 2.3709 & Associated with Sunburst \\
101.6 & 9.2 & 30.05 &  &  \\
101.8 & 19.0745 & 21.65 &  &  \\
101.11 & 26.362 & 0.5 &  &  \\
101.12 & $-$24.75 & $-$16.6285 &  &  \\
\hline
41.2 & $-$21.8885 & $-$8.1885 & 1.1860 & Foreground galaxy \\
41.1 & 15.1095 & 5.9425 &  &  \\
\hline
93.1 & 24.0365 & $-$6.8465 & 3.5100 & Background galaxy \\
93.2 & $-$35.6 & 13.646 &  &  \\
\hline
94.1 & 23.935 & $-$7.05 & 3.5100 & \dittoclosing \\
94.2 & $-$35.05 & 15.6 &  &  \\
\hline
95.1 & 23.149 & $-$9.227 & 3.5100 & \dittoclosing \\
95.2 & $-$32.65 & 22.274 &  &  \\
\hline
\hline
\end{tabular}
\end{table}

\section{Best-fit parameters for lens model}
\label{appx:lensmodelparameters}

We present the best-fit parameters of our fiducial lens model in Table \ref{tab:lens_parameters}. $\Delta_{\mathrm{RA}}$ and $\Delta_{\mathrm{DEC}}$ are measured relative to $(\mathrm{RA}_{\mathrm{BCG}}, \, \mathrm{DEC}_{\mathrm{BCG}}) = (15{:}50{:}07.115, \, -78{:}11{:}29.953)$. $e$ is the ellipticity given as $e = 1 - a/b$. $\mathrm{PA}$ is the position angle measured East of North where North is "up". Length scales $r_{\mathrm{trun}}$ and $r_{\mathrm{core}}$ are reported in units of arcsecond. Bracketed parameters were fixed and not allowed to vary in the lens model optimization. Empty fields indicate the parameter was not used. Profiles 1-4 and 6-11 (from the top) are \texttt{jaffe}, 5 is \texttt{gals}, and 12-13 are \texttt{sie} as explained in the \texttt{GLAFIC} user manual \citep{Oguri_2010}.

\begin{table*}
\renewcommand{\arraystretch}{1.2}
\caption{Best-fit lens model parameters.}
\label{tab:lens_parameters}
\begin{tabular}{p{1.2cm}|c|c|c|c|c|c|c|c}
\hline
\hline
Halo ID & $z$ & $\sigma [\mathrm{km}/\mathrm{s}]$ & $\Delta$ RA [$"$] & $\Delta$ DEC [$"$] & $e$ & $\mathrm{PA}$ [$^{\circ}$] & $r_{\mathrm{trun}}$ [$"$] & $r_{\mathrm{core}}$ [$"$] \\
\hline
\hline
DM Halo & 0.4430 & 1.264661e$+$03 & $-$2.159063e$+$00 & 1.004852e$+$00 & 1.673917e$-$01 & 7.566782e$+$01 & [2.627918e$+$02] & 9.279709e$+$00  \\
\hline
NE Subhalo & 0.4430 & 3.092102e$+$02 & $-$6.180687e$+$00 & 2.989926e$+$01 & 5.951680e$-$01 & 8.796298e$+$01 & 2.627888e$+$02 & 3.497575e$+$00  \\
\hline
BCG Subhalo & 0.4430 & 5.748313e$+$02 & $-$9.966914e$-$01 & $-$9.932597e$-$01 & 4.748962e$-$01 & 1.789992e$+$02 & 5.283780e$+$01 & 5.187953e$+$00  \\
\hline
S Subhalo & 0.4430 & 3.514352e$+$02 & $-$1.110620e$+$01 & $-$3.956538e$+$01 & 7.499064e$-$01 & 1.329332e$+$02 & [2.627918e$+$02] & 9.998911e$+$00  \\
\hline
Cluster Galaxies & 0.4430 & 1.556030e$+$02 & --- & --- & --- & --- & 1.733365e$+$02 & 1.671808e$+$00 \\
\hline
Elliptical N of N Arc & 0.4430 & 1.773400e$+$02 & [7.944870e$-$01] & [3.489488e$+$01] & 4.762955e$-$01 & 8.376578e$+$01 & 1.008816e$+$01 & 2.181358e$+$00 \\
\hline
Spiral on N Arc & 0.4430 & 1.610385e$+$02 & 8.362541e$+$00 & 3.022695e$+$01 & 8.499967e$-$01 & 1.480225e$+$02 & 6.458115e$+$00 & 2.729358e$+$00  \\
\hline
Spiral at z$=$0.5578 & 0.5578 & 1.204102e$+$02 & [8.310304e$+$00] & [1.935846e$+$01] & 6.999134e$-$01 & 1.150862e$+$02 & 2.919756e$+$00 & 2.208526e$+$00 \\
\hline
Spiral at z$=$0.7346 & 0.7346 & 1.200048e$+$02 & [6.426125e$+$00] & [1.353023e$+$01] & 6.995438e$-$01 & 6.494327e$+$01 & 4.996519e$+$00 & 4.999430e$+$00  \\
\hline
E Elliptical N of NW Arc & 0.4430 & 1.417973e$+$02 & [1.507185e$+$01] & [2.751735e$+$01] & 6.998085e$-$01 & 1.522028e$+$02 & 2.485364e$+$01 & 2.546702e$+$00  \\
\hline
W Elliptical N of NW Arc & 0.4430 & 5.000014e$+$01 & [1.816574e$+$01] & [2.570850e$+$01] & 6.875296e$-$01 & 1.789596e$+$02 & 2.437321e$+$01 & 4.998623e$+$00 \\
\hline
Perturber Near 1.9 & 0.4430 & 9.918236e$+$01 & 2.246070e$+$01 & 1.817921e$+$01 & [0.000000e$+$00] & [0.000000e$+$00] & --- & 1.966557e$+$00 \\
\hline
Perturber Near 200.8 & 0.4430 & 2.007741e$+$01 & [1.983759e$+$01] & [1.985481e$+$01] & [0.000000e$+$00] & [9.000000e$+$01] & --- & 1.311701e$-$01 \\
\hline
\hline
\end{tabular}
\end{table*}

\bibliographystyle{mnras}
\bibliography{lya_sunburstarc}

\end{document}